# What sparks the radio-loud phase of nearby quasars?

Roger Coziol,[1]* Heinz Andernach,[1] Juan Pablo Torres-Papaqui[1]
René Alberto Ortega-Minakata[2] and Froylan Moreno del Rio[1]

[1]*Departamento de Astronomía, Universidad de Guanajuato, Apartado Postal 144, 36000, Guanajuato, Guanajuato, México*
[2]*Observatório do Valongo, Universidade Federal do Rio de Janeiro, Ladeira Pedro Antonio, 43, Saúde 20080-090 Rio de Janeiro, Brazil*

**ABSTRACT**

To better constrain the hypotheses proposed to explain why only a few quasars are radio loud (RL), we compare the characteristics of 1958 nearby ($z \leq 0.3$) SDSS quasars, covered by the FIRST and NVSS radio surveys. Only 22% are RL with $\log(L_{1.4GHz}) \geq 22.5$ W Hz$^{-1}$, the majority being compact (C), weak radio sources (WRS), with $\log(L_{1.4GHz}) < 24.5$ W Hz$^{-1}$. 15% of the RL quasars have extended radio morphologies: 3% have a core and a jet (J), 2% have a core with one lobe (L), and 10% have a core with two lobes (T), the majority being powerful radio sources (PRS), with $\log(L_{1.4GHz}) \geq 24.5$ W Hz$^{-1}$. In general, RL quasars have higher bolometric luminosities and ionisation powers than radio quiet (RQ) quasars. The WRS have comparable black hole (BH) masses as the RQ quasars, but higher accretion rates or radiative efficiencies. The PRS have higher BH masses than the WRS, but comparable accretion rates or radiative efficiencies. The WRS also have higher $FWHM_{[OIII]}$ than the PRS, consistent with a coupling of the spectral characteristics of the quasars with their radio morphologies. Inspecting the SDSS images and applying a neighbour search algorithm reveal no difference between the RQ and RL quasars of their host galaxies, environments, and interaction. Our results prompt the conjecture that the phenomenon that sparks the radio-loud phase in quasars is transient, intrinsic to the AGN, and stochastic, due to the chaotic nature of the accretion process of matter onto the BHs.

## 1 INTRODUCTION

Although quasars were first discovered because of their strong emission in radio (Schmidt 1963), the majority (about 90%) are really not detected, or radio quiet (RQ), and only a few percent of the radio-loud (RL) quasars are considered as powerful radio sources (PRS), i.e., having radio luminosities well above the mean observed for the whole sample of RL quasars (Sandage 1965; Sramek & Weedman 1980; Strittmatter et al. 1980; Schmidt & Green 1983; Kellermann et al. 1989; Miller et al. 1993; Kellermann et al. 1994; Ivezić et al. 2002; Jiang et al. 2007; Rafter et al. 2011).

The physical reason why only a few quasars are RL is not well understood. This situation is most problematic, considering that within the AGN paradigm only one source of activity is assumed to be involved, namely a supermassive black hole (BH) accreting matter at the centre of their host galaxies (Lynden-Bell 1969; Rawlings & Saunders 1991).

### 1.1 The different hypotheses to explain the RQ/RL dichotomy of AGNs

Within the AGN paradigm, many hypotheses were proposed to explain the peculiar radio characteristics of quasars. For example, according to the unification model for AGNs (Antonucci 1993; Urry & Padovani 1995), the BHs at the centre of galaxies are generally encircled by a torus of gas and dust. The different optical spectra shown by AGNs are thus explained by the orientation of the torus relative to our line of sight. When we look face on to the torus, the most central regions are visible, and the spectra show broad emission lines, whereas when we look at intermediate angles, these central regions are partially obscured and we see a mixture of broad and narrow lines. Finally, if we look edge on to the torus, only the most extended regions are visible and we see only narrow emission lines.

Similarly for the radio structures, it was proposed that when viewed at intermediate angles, the radio lobes and the core are all visible, making the AGN a lobe-dominated AGN, while when viewed face on, the core is brighter than the lobes (due to relativistic beaming, also known as Doppler boosting), making the AGN a core-dominated AGN (Barthel 1989; Gopal-Krishna 1995; Kimball et al. 2011). Finally, when the AGN is viewed edge on, the ionised gas regions producing the broad emission lines become invisible and the object looks like a lobe-dominated narrow-line (type 2) radio galaxy.

However, determining the orientation angle of individual AGN is extremely difficult, and evidence favouring this model is usually based on indirect approaches (e.g., Gopal-Krishna 1995; Hoekstra et al. 1997; Urry et al. 2002; Kharb & Shastri 2004; Kimball et al. 2011). This has led many researchers in the field to question the orientation hypothesis, emphasizing, in particular, that this model seems too simplistic when applied to quasars, considering the wide range of intrinsic X-ray/UV/optical/infrared properties these objects show, and too static, ignoring that AGNs are dynamic, evolving objects (see the review by Tadhunter 2008, and references therein).

The alternative to the unification model for AGNs is the "intrinsic difference conjecture", which states that RQ and RL AGNs have different central engines. One popular hypothesis advocates that the distinctive radio characteristics of AGNs are explained by a difference in BH mass, those in RL AGNs being more massive than those in RQ AGNs (Laor 2000; Wu & Han 2001; McLure & Dunlop 2001; Gopal-Krishna et al. 2008; Richings et al. 2011). However, not all researchers in the field agree on the observational evidence (e.g., Oshlack et al. 2002; Woo & Urry 2002). Furthermore, to be coherent, this hypothesis must be interpreted within the larger context of the relation found between the BH mass and the bulge mass (or velocity dispersion) of the host galaxy (Magorrian et al. 1998; Ferrarese & Merritt 2000; Gebhardt et al. 2000; Häring & Rix 2004; Peterson et al. 2005; Gültekin et al. 2009; Graham et al. 2011). Within this context, a difference in BH mass between the RQ and RL AGNs would also imply different formation processes for the galaxy hosts (Chiaberge & Marconi 2011; Kharb et al. 2012). Consequently, we expect to find a strong correlation between the radio loudness and morphology of the host. Indeed, it is frequently reported that RL AGNs seem to be located mostly in massive elliptical galaxies (Hutchings et al. 1989; Kukula et al. 1999; Nolan et al. 2001; Lacy et al. 2002; O'Dowd et al. 2002). However, not all massive elliptical galaxies are RL, despite the fact that they are all assumed to form the same way, and in similar environments (e.g., McLure et al. 1999; Woo et al. 2005). On the other hand, Capetti et al. (2009) and Kharb et al. (2012) argued that it is not the bulge mass that counts, but how the mass is concentrated in the bulge, with RL AGNs being hosted by galaxies where the mass in the bulge is "more densely packed" than in the RQ AGNs (see also Lin & Mohr 2007). How this subtle difference depends on the formation process of the galaxies, or differences in their environments, is still an open question.

Considering the various hypotheses presented above for the intrinsic difference conjecture, researchers in the field adopted two different indirect approaches. One consists in looking for some external triggering mechanism, related to different galaxy density environments, the other consists in searching for possible differences in the accretion mechanism itself (e.g., Kuncic 1999; Vilkoviskij et al. 1999; Falcke et al. 1995; Baum et al. 1995; Daly 1995).

Following the first trend, many authors claimed that radio galaxies are located in particularly dense environments (Benitez et al. 1997; Best et al. 2005b; Croft et al. 2007; Kauffmann et al. 2008; Wylezalek et al. 2013; Ramos Almeida et al. 2013). However, this seems less obvious for broad-line AGNs (known as BLAGNs or type 1 AGNs), like most quasars (Yee & Ellingson 1993; Hutchings & Neff 1997; Wurtz et al. 1997; Stevens et al. 2010; Krumpe et al. 2012). Nor is it clear whether the possible causes proposed in environment studies to explain the apparent correlation with the radio properties of AGNs are consistent with the differences in BH masses and galaxy morphologies as reported above (see also Wurtz et al. 1997; Best et al. 2007; Falder et al. 2010; Worpel et al. 2013).

Models suggesting different accretion regimes seem to fare better on this last matter. In particular, one of the most popular of such models suggests that the accretion process in RL AGNs differs from that in the RQ AGNs because their BHs are spinning more rapidly (see the review in Véron-Cetty & Véron 2000). Unfortunately, it is very difficult to measure the spin of a BH directly (e.g., Bonson & Gallo 2016). However, this model makes many predictions that can be checked observationally. For example, in the study of Sikora et al. (2007) it is explained that the "apparent gap" between the RQ and RL AGNs is an "artefact of selection effects", created by the boundary in radio loudness between two distinct populations of AGNs, those hosted by elliptical, and those hosted by disc galaxies. This is assuming the central BHs in elliptical galaxies have larger spins than those in spiral galaxies. To legitimate this assumption, some authors argue that this is due to different merger rates (e.g., Capetti & Balmaverde 2006), claiming that minor merger episodes in late-type galaxies, specifically galaxies with small bulges and dominant spiral discs, naturally produce BHs with low angular momenta (Zhang et al. 2010).

As a matter of fact, the spin hypothesis does imply a tight relation between the formation of BHs and the formation processes of galaxies. For example, Fanidakis et al. (2011) explain that during the formation of a galaxy, hot and cold gas are added to the BH by flows, triggered by the cooling of the halo of gas, by disc instabilities, and by mergers of neighbour galaxies, all these events contributing in building up the mass and spin of the BH. They then explore how the distributions of the BH spins depends on the accretion of matter on the galaxies assuming two different accretion modes: a prolonged mode, during which the accreting gas remains in the same plane during all the accretion process, and a chaotic mode, during which the accreted gas, due to its self-gravity, fragments into multiple, randomly aligned accretion patches, producing a sequence of highly variable accretion episodes. The second mode is related to major, dry (gas-poor) mergers, where the BH growth is dominated by BH-BH mergers. This leads to a bimodal spin distribution, where high spin values only occur in galaxies with the most massive BHs, and where the power of the radio jet is strongly coupled to the spin.

However, the spin paradigm may not be the only solution. According to Broderick & Fender (2011) the age of the radio source could equally be important in explaining the RL/RQ dichotomy. For this model to work, we must assume a rapid evolution of the jet forming the extended lobes (e.g., Heinz 2002). One may then expect to observe different radio morphologies, forming a sort of "age sequence": at first an AGN forms a core, then jets appear, followed rapidly with extended lobes, that, eventually, would decrease in intensity as the jets stop injecting new matter into them (e.g, Saripalli et al. 2012). This might even imply an evolutionary connection between different radio structures, like the Fanaroff-Riley types I and II (Fanaroff & Riley 1974; Ledlow & Owen 1996; Lin et al. 2010).

On the other hands, other authors have declared the radiative efficiency of the accretion process the culprit, noting that RL AGNs produce on average more energy at any wavelength than RQ AGNs (Ghisellini 1993; della Ceca et al. 1994; Ciliegi et al. 1995; Daly 1995; Wu et al. 2002; Bian & Zhao 2003; Celotti 2005; Balmaverde et al. 2008; Koziel-Wierzbowska & Stasińska 2011). For example, according to Tchekhovskoy et al. (2010), the power output of a BH surrounded by a thin accretion disc is proportional to the square of the angular frequency of the BH, which implies that, for realistic BH spin distributions, the power output can only increase by a factor of a few tens, at most. However, when the accretion disc is thick, the beaming of the jet increases, and the power output dependence on the spin becomes steeper, producing an increase by a factor of ∼ 1000.

In a similar vein, Sikora & Begelman (2013) advanced that it is the magnetic flux threading near the BH, rather than the BH spin or Eddington ratio, that is the dominant factor in launching the powerful jets, and thus determining the radio loudness of the AGN (see also Coleman & Dopita 1992). According to this model, most AGNs are RQ because the thin accretion discs that feed them are inefficient in depositing magnetic flux close to the BH. These authors also suggest that accumulation is more likely to occur during a hot accretion (or thick disc) phase, and argue that RL quasars and powerful radio galaxies only occur when a massive, cold accretion event follows an episode of hot accretion.

### 1.2 Analysis proposed in this study

To constrain further the problem, we constructed from the Sloan Digital Sky Survey Data Release 7 (SDSS DR7; Abazajian et al. 2009) a sample of quasars with redshift $z \leq 0.3$, for which radio continuum observations are available for all of them from the NRAO VLA Sky Survey (NVSS; Condon et al. 1998), and almost all of them from the Faint Images of the Radio Sky at Twenty-centimeters survey (FIRST; Becker et al. 1995). Selecting nearby quasars allows us to detect radio structures with lower radio luminosities, and using radio images in both FIRST and NVSS we can thus construct a more complete and accurate picture of the radio morphologies of the radio-detected quasars in our sample. Also, for the undetected ones, we can obtain a better estimate of their average radio luminosity by stacking radio images from FIRST centred on their positions.

After measuring the Hβ emission line in the SDSS spectra, we kept only those quasars where the broadest component of this line has a full width at half maximum (FWHM) of at least 1000 km s⁻¹ ($FWHM_{H\beta} \geq 1000$ km s⁻¹). We imposed this criterion to favour the face-on orientation, and reduce, consequently, the uncertainties introduced by the unification model for AGNs. In principle, we would expect all our quasars to have comparable "low inclination angles" in radio, and to find no, or very few, extended radio structures. On the other hand, we might expect many of the radio sources to show evidence of Doppler boosting (e.g., Kellermann et al. 1989; Cirasuolo et al. 2003; Lu et al. 2007).

For all the quasars in our sample, the mass of the BH, $M_{BH}$, is estimated based on $FWHM_{H\beta}$. The flux in the continuum at 5100 Å is also measured, to calculate the luminosity, $L_{5100}$, which is used to estimate the bolometric luminosity, $L_{bol}$. This allows us to calculate the Eddington ratio, $\Gamma = L_{bol}/L_{Edd}$, from which we can deduce something about the accretion process in terms of the product of the accretion rate, $\dot{M} = dM/dt$, by the radiative efficiency of the accretion process, $\eta$.

To gain information about the ionisation power of the AGNs, we also measured in the optical spectrum of each quasar the flux of the [OIII]λ5007 emission line and its FWHM, $FWHM_{[OIII]}$. The luminosity of this line, $L_{[OIII]}$, is proportional to the number of ionizing photons in the extended narrow-line regions of the AGN, far from the BH, while $FWHM_{[OIII]}$ yields information about the dynamics of the gas in these regions. Adding to this information the absolute magnitude in the $i$-band, $M_i$, as given in Schneider et al. (2010), also allows us to compare the amount of energy emitted in the optical to the energy emitted in radio.

Finally, central to our analysis, we exploit the proximity of the quasars in our sample to deduce, by visual inspection of the SDSS images, information about the possible morphologies of their galaxy hosts and environments. A neighbour search algorithm is also applied to quantify the galaxy densities in these environments.

Our main statistical analysis consists in establishing how all these properties, measured in the same way, vary with the different observed radio morphologies and luminosities of the quasars, and deduce from it new clues about the phenomenon that could have triggered the radio-loud phase in these AGNs.

All the physical parameters that depend on the proper distance were calculated using the currently accepted paradigm for cosmology, namely, a cold dark-matter, dark-energy dominated Universe (LCDM), adopting the values obtained by the full-mission Planck observations of the temperature and polarization anisotropies of the CMB (Planck Collaboration 2015): $\Omega_m = 0.3089 \pm 0.0062$, $\Omega_\Lambda = 0.6911 \pm 0.0062$ and $H_0 = 67.74 \pm 0.46$ km s⁻¹ Mpc⁻¹. However, this choice is not critical, since, considering the small range in redshift covered by the quasars in our sample, different cosmologies and K corrections yield differences of the order of a few percent, which is well below the uncertainties on the parameters measured and derived in this study.

## 2 DATA

### 2.1 Selection of the sample

Our sample of quasars is drawn from the Sloan Digital Sky Survey quasar catalog produced by Schneider et al. (2010),

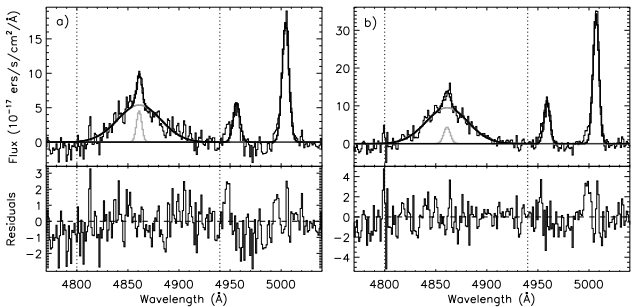

**Figure 1.** Two examples of the gaussian fitting process on two different quasars in our sample, where two components of H$\beta$, one broad and one narrow, were necessaries.

which is based on SDSS DR7, and has an astrometric accuracy of $\leq 0.1$ arcsecond. Our original sample included 2182 quasars with a redshift $z \leq 0.3$ and $FWHM_{H\beta} \geq 1000$ km s$^{-1}$. Applying this criterion assures us that the optical emission is dominated by the accretion disc, favouring AGNs with small orientation angles relative to the line of sight. This also eliminates the most face-on sources, like blazars, OVVs and BL Lacs objects, which have a spectrum dominated by the continuum. Those AGNs, however, are very rare.

To apply the FWHM criterion, first, using the maps of Schlegel et al. (1998), we corrected the SDSS DR7 spectra for Galactic reddening, adopting an average extinction law with $R_V = 3.1$ (Cardelli et al. 1989). Then, using the redshifts in the SDSS DR7 catalog, we converted them to their respective rest frames. Before measuring the emission lines, a special program written in IDL was applied to fit on each quasar spectrum a power law to its continuum over a wavelength region containing the H$\beta$ and the doublet emission lines [OIII]$\lambda$4959 and [OIII]$\lambda$5007. This was done iteratively (with more than a 1000 iterations on average), adopting the solution that minimises the rms in the residual. Each spectrum was then normalised by its proper power law, and the FWHMs and fluxes of the emission lines were measured by fitting gaussian profiles. The fits were done iteratively, with the possibility of changing the centres of the gaussians or adding (mostly for H$\beta$) more than one component between each iteration. In Fig. 1 we show two examples of fits obtained for two different quasars in our sample. From the statistics on all the residuals, we estimate a mean uncertainty on the FWHM of 3% for the broadest component of H$\beta$ and 2% for the two oxygen lines. The mean uncertainty on the fluxes is dominated by the uncertainty in calibration, which is of the order of 20% (Stoughton et al. 2002).

Eliminating the spectra that show some deficient pixels or dubious emission features we acquired a final sample of 1958 quasars ($\sim$ 90% of the original sample). In Table 1, we show a sample of the data with all the physical parameters that will be compared in our analysis. The exact meaning of these parameters will be explained in the next three subsections. In column 1 we give the SDSS DR7 quasar designation and in column 2 its SDSS redshift, $z$, as taken from Schneider et al. (2010). These are followed by the FWHM of the broadest component of H$\beta$ (col. 3), $FWHM_{H\beta}$, the luminosity of the continuum at 5100 Å, $L_{5100}$ (col. 4), the mass of the BH, $M_{BH}$ (col. 5), the bolometric luminosity, $L_{bol}$ (col. 6), the Eddington ratio, $\Gamma = L_{bol}/L_{Edd}$ (col. 7), the FWHM and luminosity of the [OIII]$\lambda$5007 emission line, $FWHM_{[OIII]}$ and $L_{[OIII]}$ (in cols. 8 and 9 respectively). The results for the galaxy density in the environment of all the quasars, parameterised as the number $N_R^{-19}$, are given in columns 10, 11 and 12, for the respective radii $R = 0.5$, $R = 1.0$ and $R = 1.5$ Mpc. In columns 13, 14, 15 and 16, one can find, respectively, our classification of the morphological type of radio emission, the monochromatic 1.4-GHz radio luminosity, $L_{1.4GHz}$, or upper limits for the radio-undetected (RU) quasars, the largest (projected) linear size, LLS, for the extended radio structures, and a flag stating how the LLS was determined. The last column (col. 17) indicates the results of our visual inspection of the SDSS images: r for optically resolved, e for elliptical or s for spiral, i for possible merger or interaction, and b if it is the brightest object in a large scale structure. In the next three subsections, we explain how all these parameters were obtained.

## 2.2 Radio morphologies and luminosities

We first cross-correlated the positions of the selected quasars with entries in the FIRST and NVSS catalogs within a radius of 2 and 6 arcseconds, respectively. This allows us to detect all the radio sources with 1.4-GHz flux $\geq 1$ mJy in FIRST and $\geq 2$ mJy in NVSS. Next we visually inspected all FIRST and NVSS images at least 1 Mpc on a side at the QSO redshift, in order to search for (a) core radio emission fainter than the FIRST and NVSS catalog thresholds, and (b) extended radio structures associated with the QSOs, regardless of the presence of a radio core (e.g. Lu et al. 2007). We made sure that these extended structures are: 1) not associated with another optical source or infrared source in the field (Rafter et al. 2011), using the VizieR catalog access tool at CDS (Ochsenbein et al. 2002), and that 2) most of the sources, except 5 (classified as X in Table 1), are relatively well aligned with the optical galaxy at the position of the radio core. We did find a few quasars with extended lobes in NVSS and weak core in FIRST, but, despite specifically looking for them, these cases turned out to be rare. This may be due to both, the good sensitivity of FIRST to radio cores, and to our rather small sample of only low-redshift quasars.

During our visual inspection we noted many sources fainter than the NVSS and FIRST catalog thresholds that were clearly coincident with the QSOs. We thus included all such sources in FIRST within 2 arcsec and stronger than 0.5 mJy, and within 6 arcsec in NVSS if stronger than 1.0 mJy. These limits are equivalent to 3 times the noise level of the respective survey images. The fluxes were determined either by gauss-fitting for point sources or by image integration if they appeared extended. We classified all QSOs with associated radio sources as radio detected (RD) and all others as radio undetected (RU). Note that 70 quasars in our sample are not covered by the FIRST survey, and an upper limit of 1.0 mJy (from NVSS) was used for these. In total, we classified $\sim$ 22% (431) of the quasars in our spectroscopic sample as RD. The ratio RD/RU of quasars in our sample is consistent with the ratio RL/RQ typically found for quasars at low redshift (e.g., Jiang et al. 2007; Kimball et al. 2011).

To verify the reliability of the detections below the FIRST catalog limit of 1.0 mJy, we estimated the number of

**Table 1.** Example of table for data of the 1958 quasars in our sample

| (1) SDSS J Name | (2) z | (3) FWHM Hβ (km s⁻¹) | (4) L₅₁₀₀ (log) (W) | (5) M_BH M_☉ | (6) L_bol (log) (W) | (7) Γ | (8) FWHM [OIII] (km s⁻¹) | (9) L_[OIII] (log) (W) | (10) N⁻¹⁹_{0.5} | (11) N⁻¹⁹_{1.0} | (12) N⁻¹⁹_{1.5} | (13) Radio class. | (14) L_{1.4GHz} (log) (W Hz⁻¹) | (15) LLS (kpc) | (16) LLS flag | (17) Vis class |
|---|---|---|---|---|---|---|---|---|---|---|---|---|---|---|---|---|
| 000102.18 − 102326.9 | 0.2943 | 6319 | 37.55 | 8.60 | 38.50 | 0.06 | 366 | 35.56 | | | | RU | < 23.11 | | | ...b |
| 073115.65 + 423944.5 | 0.2007 | 2251 | 37.19 | 7.47 | 38.15 | 0.33 | 367 | 34.51 | 2 | 5 | 10 | RU | < 22.75 | | | rs.. |
| 073309.20 + 455506.2 | 0.1414 | 3568 | 37.84 | 8.29 | 38.79 | 0.23 | 373 | 35.37 | 0 | 2 | 4 | C | 23.07 | 19 | F | ..i. |
| 074906.50 + 451033.8 | 0.1921 | 4160 | 37.74 | 8.35 | 38.69 | 0.15 | 399 | 35.85 | 0 | 2 | 4 | T | 25.23 | 582 | M | .... |
| 091133.85 + 442250.0 | 0.2976 | 4324 | 37.49 | 8.23 | 38.45 | 0.11 | 581 | 35.25 | 1 | 2 | 7 | J | 26.07 | 93 | M | re.b |
| 093347.76 + 211436.4 | 0.1722 | 4288 | 37.28 | 8.09 | 38.23 | 0.10 | 531 | 35.10 | 4 | 6 | 10 | L | 24.06 | 431 | M | rei. |
| 105609.79 + 551604.0 | 0.2563 | 3007 | 37.46 | 7.89 | 38.41 | 0.07 | 442 | 34.65 | 2 | 3 | 6 | X | 23.66 | 404 | M | rsi. |

A lack of data in cols. 10, 11 and 12 denote quasars that were not included in the neighbour galaxy search (see explanation in the text).
For RU quasars col. 14 gives an upper limit for the radio luminosity based on a flux of 0.5 mJy if covered by FIRST and 1.0 mJy if only covered by NVSS.
No linear sizes are given in col. 15 for unresolved radio sources.
Table 1 is published in its entirety in the electronic edition. A portion is shown here for guidance regarding its form and content.
See Section 2.1 for a full description of the columns.

random coincidences from three independent source counts deeper than FIRST, namely, Bondi et al. (2008); Ibar et al. (2009); Vernstrom et al. (2015). These surveys found the number of sources with fluxes between 0.5 and 1.0 mJy, to be between 50 and 80 deg⁻², and about twice that number for *all* sources stronger than 0.5 mJy. This implies that within a radius of 2 arcsec around the 1958 QSOs of our sample, the random expectation is to find 0.15 sources between 0.5 and 1.0 mJy, and 0.3 sources above 0.5 mJy. In fact, we found 77 and 431 such sources, respectively, which implies that none of these sources is expected to be a random coincidence. This procedure was also followed by de Vries, Becker & White (2006).

Adopting the higher of the two flux densities, 18% from FIRST and the rest from NVSS, the radio luminosities of the RD quasars at 1.4 GHz were calculated using the relation (Weedman 1988):

$$L_{1.4 \text{ GHz}} = 4\pi \; d_p^2(z)(1 + z)\left(\frac{f_{1.4 \text{ GHz}}}{\text{Jy}}\right) \text{ W Hz}^{-1} \qquad (1)$$

Here $d_p(z)$ is the proper distance at the corresponding redshift, as calculated numerically following Wright (2006). To this equation, we added a K-correction term, $L_{\nu_e} = L_{\nu_o}(\nu_0/\nu_e)^{\alpha} = L_{\nu_o}(1 + z)^{-\alpha}$ (Weedman 1988), where we adopted a radio spectral index of $\alpha = -0.5$ typical for QSOs (Kellermann et al. 1989). The mean uncertainty on the radio luminosities is estimated to be 11%. This was determined by taking into account the flux density error, $\sigma_n = \text{rms} \times \sqrt{N_b}$, where $N_b$ is the integration area in units of the beam area. The beam area is defined as $\Omega_b = 1.13\,\theta_b^2$, the beam widths being $\theta_b = 5.4$ arcseconds for FIRST and 0.75 arcmin for NVSS (White et al. 1997; Condon et al. 1998).

To determine an average radio flux for the RU quasars, we stacked 1466 FIRST images (61 RU quasars were not covered by FIRST), each image covering a region of 1′ × 1′ on the quasar position. The stacking process consists in calculating the arithmetic mean pixel of the stacked images. We confirmed that standard error decreases as the square root of the number of images stacked, $\sqrt{N}$, such that the stacked image, shown in Fig. 2, has a 3-$\sigma$ detection limit of 0.012 mJy beam⁻¹. In this image we measured a flux for the central source of 0.15 mJy with a signal-to-noise-ratio $S/N \sim 40$. At the mean redshift of the RU quasars in our sample, $\bar{z} = 0.241$, this corresponds to a mean radio luminosity of $L_{1.4\text{GHz}} = 2.5 \times 10^{22}$ W Hz⁻¹, which is between one and 2.5 times times lower than the radio luminosity that separates the majority of spiral from elliptical galaxies (e.g., Ekers & Kotanyi 1978; Condon 1988).

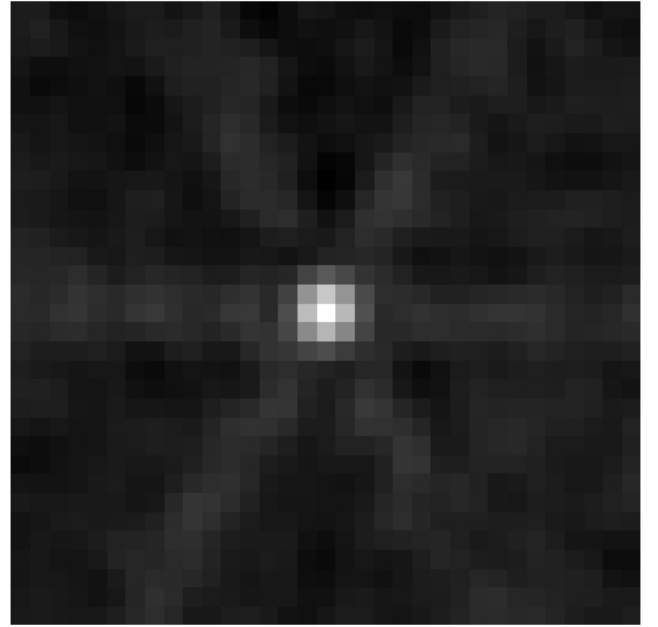

**Figure 2.** Arithmetic mean obtained by stacking the FIRST images of 1466 RU quasars. Each image covers an area of 1′ × 1′ centered on the position of the quasar. The radial structure is an artifact produced by the limited uv coverage of the FIRST survey.

Following our examination of the FIRST and NVSS images, we classified the radio morphologies of the RD quasars adopting the types defined by Kimball et al. (2011). Our whole sample of RD quasars can be separated in four basic types: 366 quasars are compact (C), 11 show a core with a jet (J), 8 have a core and lobe (L) and 41 are triple (T), showing a core with a jet and two lobes. Only five quasars could not be classified according to this simple framework (identified as X in Table 1). An example of each radio structure is shown in Fig. 3. The majority (84.7%) of the RD quasars in our sample are C-type, and only 9.5% (2.1% of the whole sample) have a T classification.

In Fig. 4, we show the distribution of the radio luminosity of the RD quasars as a function of redshift. The three different curves correspond to the three flux limits applied as selection criteria: from top to bottom, (dashed) 2 mJy for the NVSS catalog sources, (dotted) 1 mJy for the FIRST cata-log sources and NVSS sources coinciding with the positions of the quasars, (solid) 0.5 mJy for sources found to coincide with the quasar positions in the FIRST images. Due to the relatively narrow redshift range of our sample, the lower limit in luminosity, from the lowest to the highest $z$, only increases by a factor $(0.3/0.07)^2 \sim 16$. Therefore, we would expect our detection of radio structures to be marginally affected by this small increase in minimum detectable radio luminosity. One way how to estimate the level of completeness of our survey is to compare the average flux measured for the RU quasars, using the stacking method, with the predictions based on Figure 1 in de Vries, Becker & White (2006), where the percentage of quasars with a detected core is traced as a function of the core flux density. Based on this diagram, the probability to detect a quasar above 0.15 mJy is 18%, while we found 22% in total in our sample. This suggests that our level of incompleteness cannot be higher than a few percent.

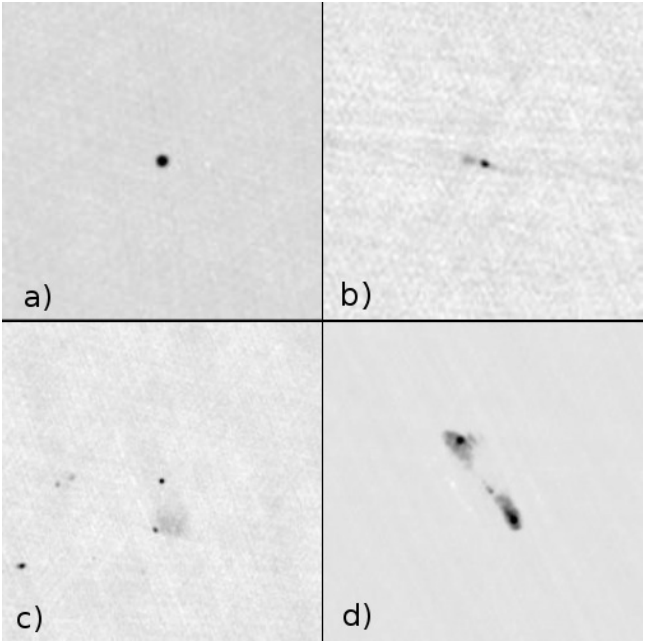

**Figure 3.** Examples of the four types of radio morphology used in our classification: a) compact (C), b) core with jet (J), c) core with one lobe (L) and d) triple: a core with jet and two lobes (T).

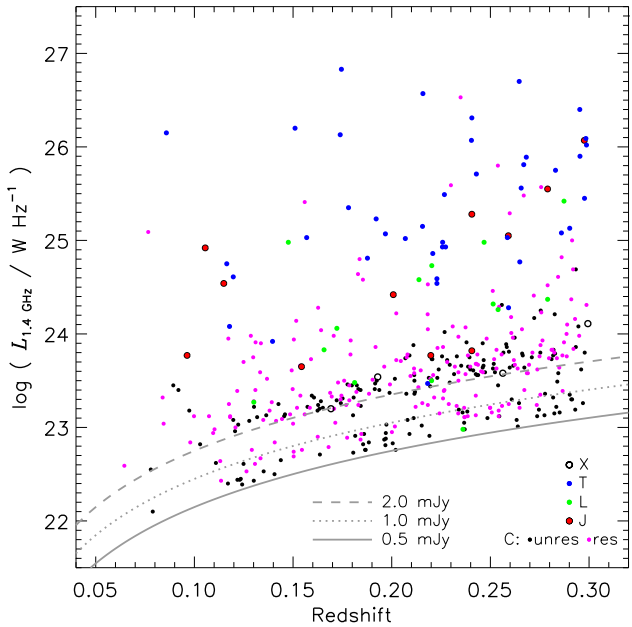

**Figure 4.** The 1.4 GHz radio luminosity as a function of the redshift for the RD quasars in our sample. The three curves correspond to the three flux limits applied as selection criteria (see Section 2.2). The different colours correspond to the different radio structures.

Increasing the area to search for extended radio structures raises the likelihood of mis-identifications (Condon et al. 1998). However, comparing with other studies, there is no evidence that our results for the extended radio structures include false associations. For example, Kimball et al. (2011) inspected areas of $4' \times 4'$ only for radio sources suspected to be more extended than $1'$, and found $619/4714 \sim 13\%$ quasars with a T-type, which is slightly higher than the $41/426 \sim 10\%$ in our survey, where we systematically inspected larger areas around all the sources. On the other hand, according to de Vries, Becker & White (2006) only 2% of the radio sources detected in their survey were found to be extended, consistent with Fanaroff-Riley type II (FRII). This percentage is significantly smaller that what we observed, if one assumes all the 41 T quasars in our sample, $\sim 10\%$ of the extended sources, are consistent with FRII (not considering their radio luminosities). Note however that in the study of de Vries, Becker & White (2006), no upper limit in redshift was applied. Moreover, according to Lu et al. (2007), taking into account selection effects in the SDSS radio quasars sample, extended radio sources are much more common than what de Vries, Becker & White (2006) have determined. According to Fig. 3 in Lu et al. (2007), we would easily expect 19% such objects at redshift $z \leq 0.3$, which qualifies our result as conservative. These different comparisons suggest that careful selection criteria, as we applied in our study, reduced significantly the probability of false associations (further arguments supporting this claim will be given in Section 2.3).

For the 60 RD extended quasars in our sample (L, J and T types), we determined the largest angular size (LAS) of the radio structures. In Fig. 5, we show an example of how the LAS was measured. Whenever the source presented hot spots at its two extremities, FIRST was used to determine the LAS. For the extended sources that are resolved out in FIRST or that have very low surface brightness, the NVSS images were used to measure the LAS. The largest linear size (LLS) was then calculated using the relation described in Eq. 2 (Weedman 1988), with a mean uncertainty of 19%:

$$\text{LLS} = 1.163 \left( \frac{\text{LAS}}{\text{arcmin}} \right) \left( \frac{d_p}{1+z} \right) \text{ Mpc} \qquad (2)$$

For the RD quasars classified as C, which are resolved according to FIRST (56% of the C quasars), we adopt an LAS equal to the deconvolved source sizes as listed in the FIRST catalog.

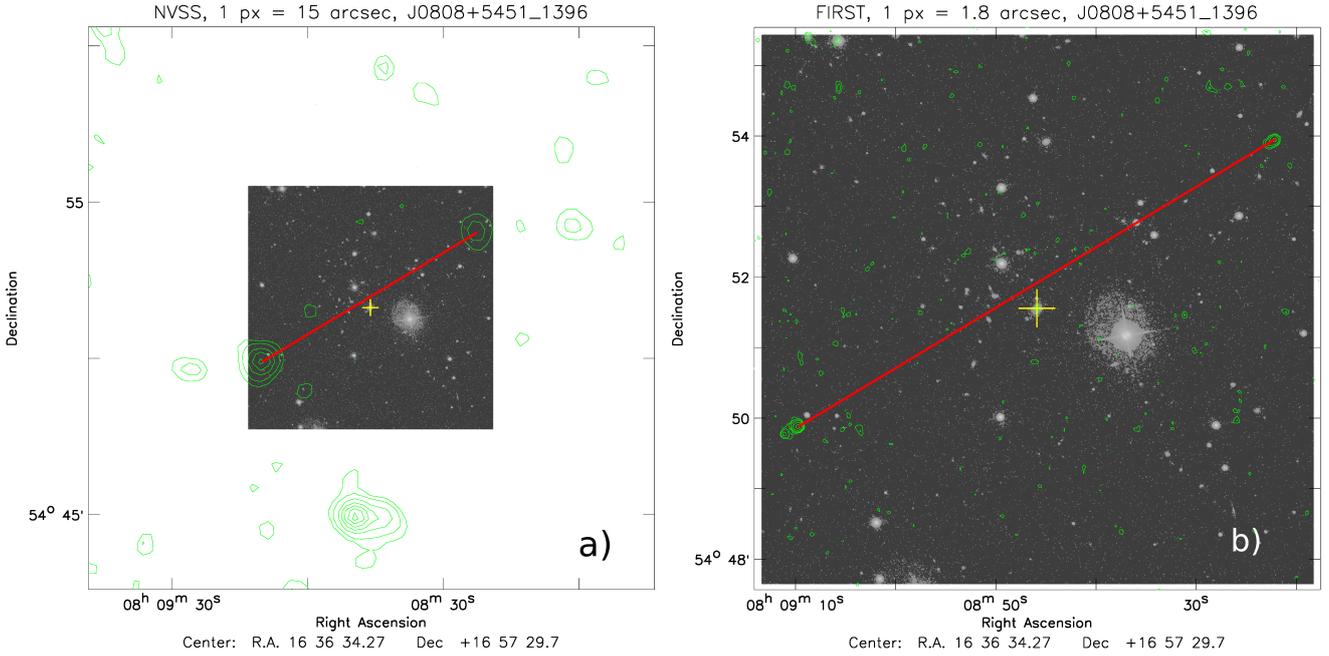

**Figure 5.** Illustration how the largest angular size (LAS) of the radio structure was measured: radio contours of a) NVSS images, b) FIRST. Both are overlays on SDSS *r*-band images. The LAS is defined as the angular distance between detectable radio emission features furthest away and on opposite sides of the host.

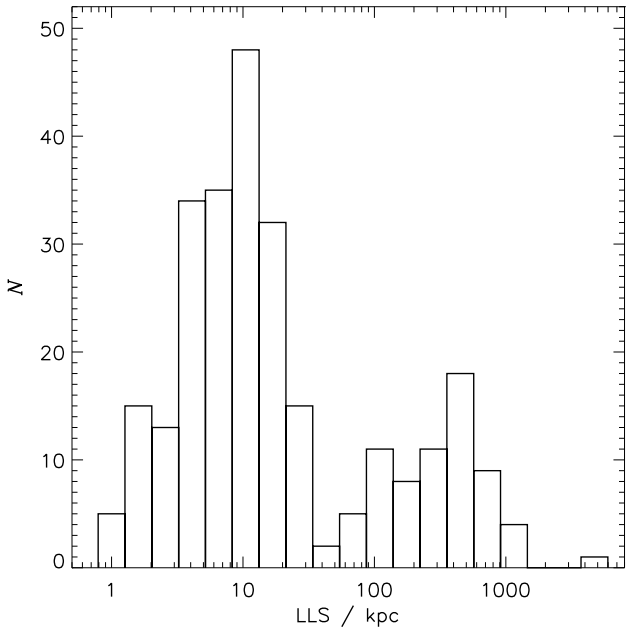

**Figure 6.** Distribution of the linear sizes of the 266 resolved radio sources in our sample. One quasar, J0931+3204, have a LLS exceeding 4 Mpc.

In Fig. 6 we show the distribution of the LLS for all the "resolved" RD. The distribution is apparently bimodal. The 206 resolved, C-type quasars have a median LSS of only 8 kpc (a mean of 11 kpc), while the 60 extended radio structures have a median of 348 kpc, which is about twice the typical value for radio galaxies (e.g., Best et al. 2005a; Lin et al. 2010). However, we cannot be certain that this bimodality is genuine. This is because although we are sure that we did not miss any extended source with an intermediate LLS that might fill the gap between the C and T quasars, we cannot discard the possibility of C-type quasars with extended structures within this range, but near the survey limit. Indeed, we note that of the 366 C quasars in our sample, 110 have a ratio $f_{NVSS}/f_{FIRST} \geq 1.2$, implying that they might have extended structures on scales between the FIRST and NVSS angular resolutions. This suggests that the bimodality in the LSS radio structures can be due, in part, to an observational bias.

### 2.3 Giant Radio Quasars in our Sample

Whenever the radio emission of an extended source exceeds a projected linear size of 1 Mpc, the source is usually called a giant radio galaxy (GRG) or quasar (GRQ). Although GRQs exist, they are nonetheless rare (de Vries, Becker & White 2006; Lu et al. 2007; Rafter et al. 2011). Previous examples in the literature can be found in Buchalter et al. (1998), Bhatnagar, Gopal-Krishna & Wisotzki (1998), and Singal, Konar & Saikia (2004). The largest GRQ claimed to date is FBQS J0204−0944, which at a redshift $z = 1.004$ has a LLS = 3.0 Mpc (Kuźmicz & Jamrozy 2012).

Remarkably, three quasars in our sample qualify as GRQ. Two of them (J0843+2037 and J1435+4948) have a LLS of just over 1 Mpc, and a third one (J0931+3204) even has a LLS > 4 Mpc. In Fig. 7 we reproduce the NVSS radio contours and show the SDSS colour images of their respective hosts. A brief description of these extended radio sources follows.

The GRQ J0843+2037 is hosted by SDSS J084347.84+203752.4 at a redshift $z = 0.2276$. While this GRQ was reported by Condon et al. (2013), only our visual inspection of the NVSS image led us to measure an angular size of 5.38′ for this source, corresponding to an LLS of 1.27 Mpc. Its radio morphology is that of an FR II, with a prominent bridge connecting the radio core with the NW hotspot (see Fig. 7a), while the SE lobe is amorphous in NVSS, indicating a plume due W of the SE hotspot in FIRST (not shown here). The NVSS-to-FIRST integrated flux ratio for the SE lobe is 25.8 mJy/7.8 mJy = 3.3, which indicates that this source is dominated by diffuse emission. In the FIRST image the source is rather symmetric with an armlength ratio (NW/SE) of 1.04, and a misalignment angle ∼ 5.2° between the core and the FIRST hotspots. There is an optical/IR object close to the SE hotspot, the galaxy UGCS J084356.20+203557.1 in UKIDSS-DR9 (Lawrence et al. 2007). However, located at ∼ 1.7″ SE of the FIRST peak position, it is unlikely to be the host of the SE lobe. The NW hotspot has no optical/IR counterpart within 7″ of its FIRST peak position.

The GRQ J1435+4948 lies at a redshift $z = 0.1661$ and is hosted by the quasar SBS 1433+500 (alias LEDA 2353978 and CSO 670). It also coincides with the X-ray source 1RXS J143509.6+494814. While the NVSS contour map (Fig. 7b) suggests a one-sided emission due NW, FIRST resolves this source into the QSO core and a compact, 7.1-mJy radio source associated with SDSS J143506.93+494836.6, at a photometric redshift $z_{ph} = 0.479$. This radio source was wrongly interpreted by Kimball et al. (2011) as a single lobe, associated to the quasar SBS 1433+500. Discarding this one-sided radio lobe, we interpret the two diffuse sources in the NVSS image, symmetrically spaced along PA∼ 155° on opposite sides of the QSO, as relic lobes of an FR II radio source. The lobes are completely resolved out in FIRST, consistent with genuinely diffuse radio emission from aged radio lobes, for which the hotspots have faded away, being now barely distinguishable at the 0.6 mJy level in the FIRST image. The radio source ∼ 1.8′ WSW of the SE of these diffuse lobes is due to the unrelated spiral galaxy 2MASX J14350419+4945538. With a LAS of 5.7′ between the outer edges of the relic lobes, GRQ J1435+4948 has a LLS of 1.00 Mpc. We estimate its total flux density from the NVSS image as 35 mJy, corresponding to $\log(L_{1.4GHz}) \sim 24.42$ W Hz$^{-1}$.

The GRQ J0931+3204 is hosted by the quasar 2MASX J09313900+3204006 at a redshift $z = 0.2257$. The NVSS contour image (Fig. 7c) shows that it has an asymmetric FR II morphology with a 19.9′ projected size. From the FIRST image (not shown here) we measure the longer NE arm to have 11.9′ at PA = 61.2° and the shorter SW arm to have 7.87′ at PA = 243.8°. This corresponds to a misalignment angle of only 2.6° over a total projected linear size of 4.45 Mpc. Both outer lobes are dominated by extended emission, with NVSS-to-FIRST integrated flux ratios of 23.7 mJy/13.4 mJy = 1.77 for the SW lobe and 17.9 mJy/6.1 mJy = 2.9 for the NE lobe. The compact radio core coincides with the quasar, and has a NVSS-to-FIRST flux ratio of 10.4 mJy/9.4 mJy = 1.1. Its deconvolved size of

1.7″ in the FIRST survey at PA = 81° is consistent with a weak contribution from a radio jet pointing within 20° of the source's major axis. The NVSS source ∼ 2′ SSE of the NE hotspot is an unrelated double source most likely coinciding with SDSS J093230.72+320703.4 at an estimated redshift of 0.266 (Alam et al. 2015) or 0.1758 (Brescia et al. 2014).

2MASX J09313900+3204006 was first proposed as a GRG by Andernach et al. (2012), and the more complex NE lobe has recently been observed by us with the Jansky Very Large Array in D configuration at 5.5 GHz, confirming that this lobe consists of two complex hotspots (Andernach et al., in preparation). Neither of the latter, nor the SW hotspot has an optical identification in SDSS. Moreover, in the FIRST image the SW hotspot clearly shows a cometary shape with the peak brightness at the outermost (SW) edge and fainter, extended emission pointing towards the quasar host, which is better seen in the NVSS image. Its LLS of 4.45 Mpc makes J0931+3204 the largest GRQ currently known, and the forth-largest of all known GRGs, only exceeded by [MKJ2008] J1420−0545 (Machalski et al. 2008), J1234+5318 (Banfield et al. 2015), for which a spectroscopic redshift has recently been obtained by us with the 10.4 m Gran Telescopio Canarias (GTC; Andernach et al., in preparation), and 3C 236, the very first example of a GRG discovered by Willis, Strom & Wilson (1974).

Based on the radio morphology of J0931+3204, we can estimate the probability of having a source by chance in two (small) cones on opposite sides of the central source. The radius vectors from the core to the NE and SW lobes are misaligned by 2.6°. The SW lobe of ∼ 18 mJy lies at ∼ 8′ distance from the host quasars, and the NE lobe of ∼ 24 mJy at 12′. If we assume an opening angle of 10° on each side, which is four times the misalignment angle, then we seek the probability of having an 18-mJy source in a cone of 10° and 12′ long, together with a 24-mJy source over the same cone, but only 8′ long. In NVSS there are 10.7 deg$^{-2}$ sources stronger than 18 mJy and 8.3 deg$^{-2}$ sources above 24 mJy, which yields chance probabilities of $3.2 \times 10^{-4}$ and $1.7 \times 10^{-4}$, respectively, for the NE and SW lobes. For the probability that this occurs at the same time, i.e., having two sources in the sum of the area ($4.9 \times 10^{-5}$ deg$^2$), Poisson statistics yields the probability $p(N) = (n^N/N!)e^{-n} \sim p(2) = ((5.3 \times 10^{-4})^2/2)e^{-5.3 \times 10^{-4}} \simeq 1.4 \times 10^{-7}$.

Note that this probability would also apply for two adjacent cones (or a single one of 20° opening) on one side of the host, or to any two cones with any angle between them, or even a square of the same area. Thus, for an aligned source as J0931+3204 the probability should be reduced by another factor of, say, 20°/360° = 1/18. So the probability of such an alignment to occur by chance in NVSS is ∼ $7.8 \times 10^{-9}$. While this is already a very small probability, it still does not take into account the (less quantifiable) characteristics of that source, namely (a) the cone or cometary shape of the SW hotspot, (b) the absence of any optical counterpart for either hotspot, and (c) the orientation of the radio core within 20° of the source major axis. We thus consider the GRQ nature of J0931+3204 as quite solid.

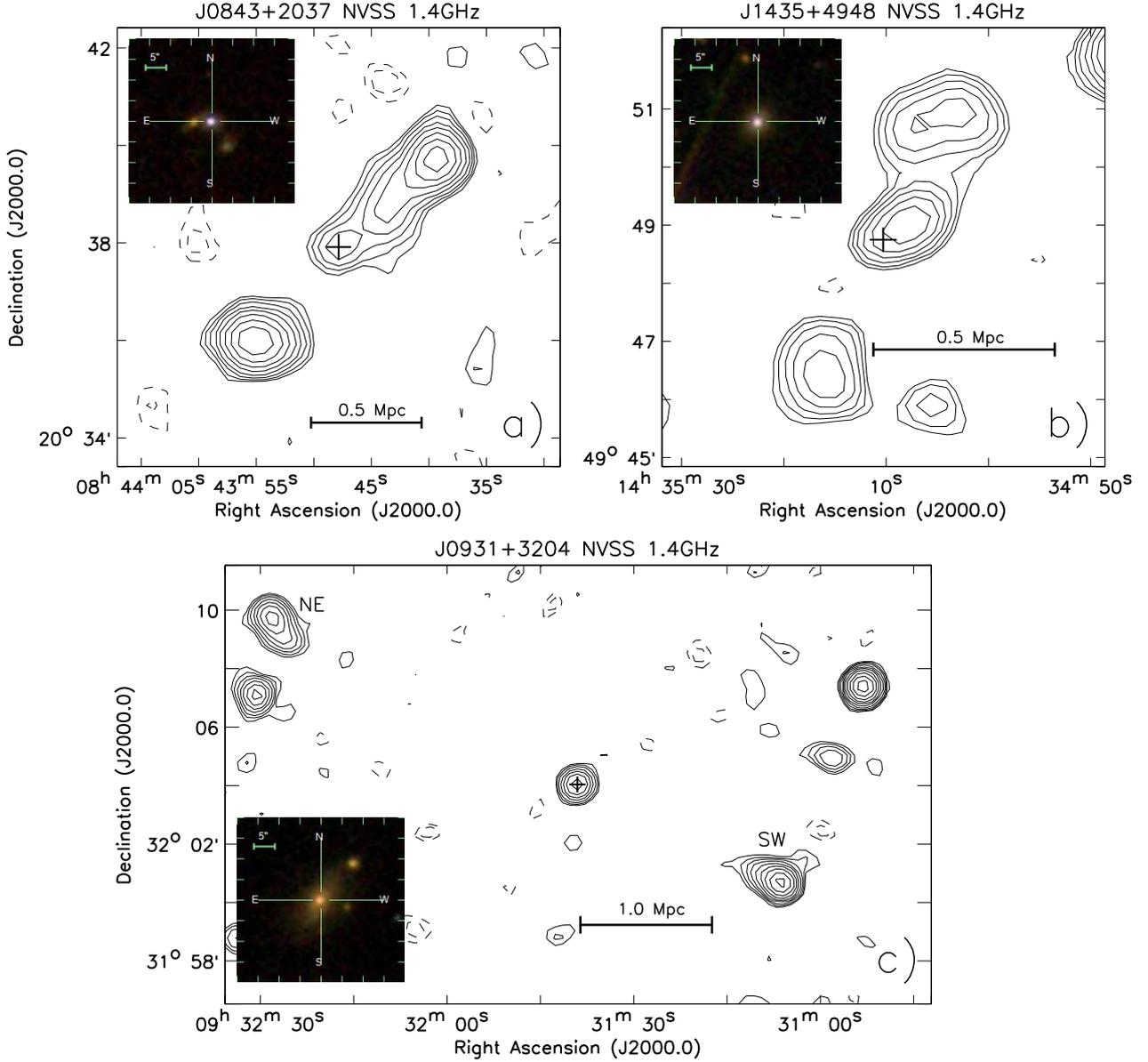

**Figure 7.** Contour plots of the NVSS images for the three giant radio quasars found in the present paper. (a) J0843+2037 at $z = 0.2276$, (b) J1435+4948 at $z = 0.1661$, and (c) J0931+3204 at $z = 0.2257$. For the latter source the NE and SW lobes are labelled for clarity. Contours are plotted at levels of $\pm 1 \times 2^{n/2}$ mJy beam$^{-1}$ for $n = 0, 1, 2, \ldots$ with negative levels plotted with dashed lines. The half power beam width of the NVSS is 45″ and it has an average rms noise level of 0.45 mJy beam$^{-1}$. Central plus signs indicate the location of the respective host quasar. The insets show a $g, r, i$ composite image from SDSS, covering an area of $40'' \times 40''$, and centered on the host quasar. See sect. 2.3 for details of these sources.

## 2.4 BH masses, Eddington ratios and [OIII]$\lambda$5007 luminosities

The mass of the BH was obtained by applying the virial theorem $2\langle T \rangle = \langle V \rangle$, where $T$ is the kinetic energy of the gas in the broad-line region (BLR) and $V$ is the gravitational potential energy. Assuming a non-relativistic potential:

$$M_{\mathrm{BH}} = \frac{R_{\mathrm{BLR}} v^2}{G} \quad (3)$$

where $R_{\mathrm{BLR}}$ is the radius of the broad-line region, $v$ is the mean velocity of the gas in the broad-line region and $G$ is the gravitational constant. For the mean velocity of the gas we used $FWHM_{\mathrm{H}\beta}$. Assuming equipartition of energy, the velocity is approximated by the equation (Blandford et al. 1990):

$$v = \frac{\sqrt{3}}{2} FWHM_{\mathrm{H}\beta} \quad (4)$$

To determine the radius of the broad-line region, we

**Table 2.** Results of the eye examination of the quasar SDSS images

| (1) Radio type | (2) Total | (3) PL frac. (%) | (4) E frac. (%) | (5) M/I (%) | (6) Is (%) | (7) BG (%) | (8) Sp frac. (%) | (9) M/I (%) | (10) Is (%) | (11) BG (%) |
|---|---|---|---|---|---|---|---|---|---|---|
| RU | 1527 | 39 |  | 16 | 31 | 5 |  | 16 | 31 | 5 |
|  |  |  | 59 | 18 | 21 | 15 | 41 | 17 | 28 | 7 |
| C | 366 | 41 |  | 32 | 41 | 30 |  | 32 | 41 | 30 |
|  |  |  | 53 | 35 | 30 | 20 | 47 | 24 | 34 | 22 |
| J | 11 | 9 |  | 100 | 0 | 100 |  | 100 | 0 | 100 |
|  |  |  | 90 | 44 | 22 | 33 | 10 | 100 | 0 | 100 |
| L | 8 | 33 |  | 33 | 33 | 33 |  | 33 | 33 | 33 |
|  |  |  | 100 | 50 | 33 | 0 | 0 | 0 | 0 | 0 |
| T | 41 | 46 |  | 16 | 16 | 16 |  | 16 | 21 | 16 |
|  |  |  | 64 | 0 | 21 | 50 | 36 | 12 | 62 | 12 |

For comparison sake, the percentage of M/I, Is and BG in the PL are included above the values for the E and repeated for the Sp.

used the empirical relation established by Greene & Ho (2005):

$$R_{\mathrm{BLR}} = (30.2 \pm 1.4) \left( \frac{\lambda L_{5100}}{10^{37}\ \mathrm{W}} \right)^{0.64 \pm 0.02} \mathrm{lightdays} \quad (5)$$

where $L_{5100}$ is the monochromatic luminosity of the continuum at $\lambda = 5100$ Å, which we measured in the SDSS spectrum. Substituting $R_{\mathrm{BLR}}$ in the previous equation for the mass of the BH we get:

$$M_{\mathrm{BH}} = (4.4 \pm 0.2) \times 10^6 \left( \frac{\lambda L_{5100}}{10^{37}\ \mathrm{W}} \right)^{0.64 \pm 0.02} \left( \frac{FWHM_{\mathrm{H}\beta}}{10^3\ \mathrm{km\ s^{-1}}} \right)^2 \quad (6)$$

Based on the errors obtained by Greene & Ho (2005) for their empirical relation, the mean uncertainty on the BH masses is estimated to be of the order of 6%.

The monochromatic luminosity at $\lambda = 5100$ Å is calculated using the formula (Weedman 1988):

$$L_{\lambda} = 4\pi\ d_p^2(z)(1+z)^3 f_{\lambda 0} \mathrm{W\ m^{-1}} \quad (7)$$

where $f_{\lambda 0}$ is the observed monochromatic flux in the continuum. Then, the K correction added to this equation takes the form $\mathrm{L}_{\lambda e} = \mathrm{L}_{\lambda o} (1 + z)^{2+\alpha}$, using the optical spectral index $\alpha = -0.5$ typical for quasars (Weedman 1988; Osterbrock & Ferland 2006).

The luminosity of the [OIII]$\lambda 5007$ line was calculated similarly, using:

$$L = 4\pi d_p^2 (1 + z)^2 f_0 \quad (8)$$

where $f_0$ is the observed flux obtained by integrating the gaussian profile fitted to the line. The uncertainties on the luminosities are dominated by the calibration in flux, and are consequently of the order of 20%.

The Eddington ratio, $\Gamma = L_{\mathrm{bol}}/L_{\mathrm{Edd}}$, was calculated using for the bolometric the approximation proposed by Kaspi et al. (2000):

$$L_{\mathrm{bol}} \simeq 9 \times \lambda L_{5100} \quad (9)$$

Since, by definition (e.g., Frank, King & Raine 1992), the Eddington luminosity is proportional to the mass of the BH, $L_{\mathrm{Edd}} \simeq 10^{31}(M_{\mathrm{BH}}/M_\odot)$ W, then $\Gamma$ is simply a normalization of the energy produced by the mass accretion onto the BH by the mass of this BH. This is an important distinction to take into account, because for the same mass accretion rates and radiative efficiencies, massive BHs will show lower $\Gamma$ than smaller ones.

An alternative expression for the Eddington luminosity would be (King & Pringle 2006):

$$L_{\mathrm{Edd}} = \frac{M_{\mathrm{BH}} c^2}{t_{\mathrm{Edd}}} \quad (10)$$

where $t_{\mathrm{Edd}} = 4.5 \times 10^8$ yrs, which, coincidentally, is of the same order of magnitude as the timescale for extended radio structures to develop in RL quasars. Using $L_{\mathrm{Edd}}$ under this form, therefore, could be useful to compare how fast the BHs in RQ and RL quasars increased their masses (e.g. King & Pringle 2006).

### 2.5 Quasar environments and morphological types of the host galaxies

Images from SDSS centred on the position of the quasars and covering an area of at least 1 Mpc$^2$ were dowloaded for the whole sample of 1958 objects. Examining these images allows us to identify the large-scale structures where the quasars are located, either clusters or groups, or to identify isolated quasars (identified as Is in Table 2). For the quasars that are part of large-scale structures, we are also interested to know if their galaxy hosts are the brightest, or among the brightest galaxies of their structure ( see Coziol et al. 2009). These are classified as BG, in Table 2.

According to the "resolved" flag in the Sloan Digital Sky Survey quasar catalog of Schneider et al. (2010), about 60% of the quasars in our sample are optically resolved. For these quasars we estimated the morphological types of their host galaxies by eye. One of the most obvious characteristics of the resolved images, easy to recognize by eye, are extended morphological structures, consistent with the presence of spiral discs. We classified these quasars as spiral-like (Sp). We then classified the rest as early-type (E), which means that we consider them to be either elliptical or early-type spirals, since we cannot discriminate these types by eye. Our examination of the images can also reveal possible evidence of mergers or interaction (M/I). One example of each of the eight different categories used to classify the images of the quasars can be found in Fig. 8 and Fig. 9.

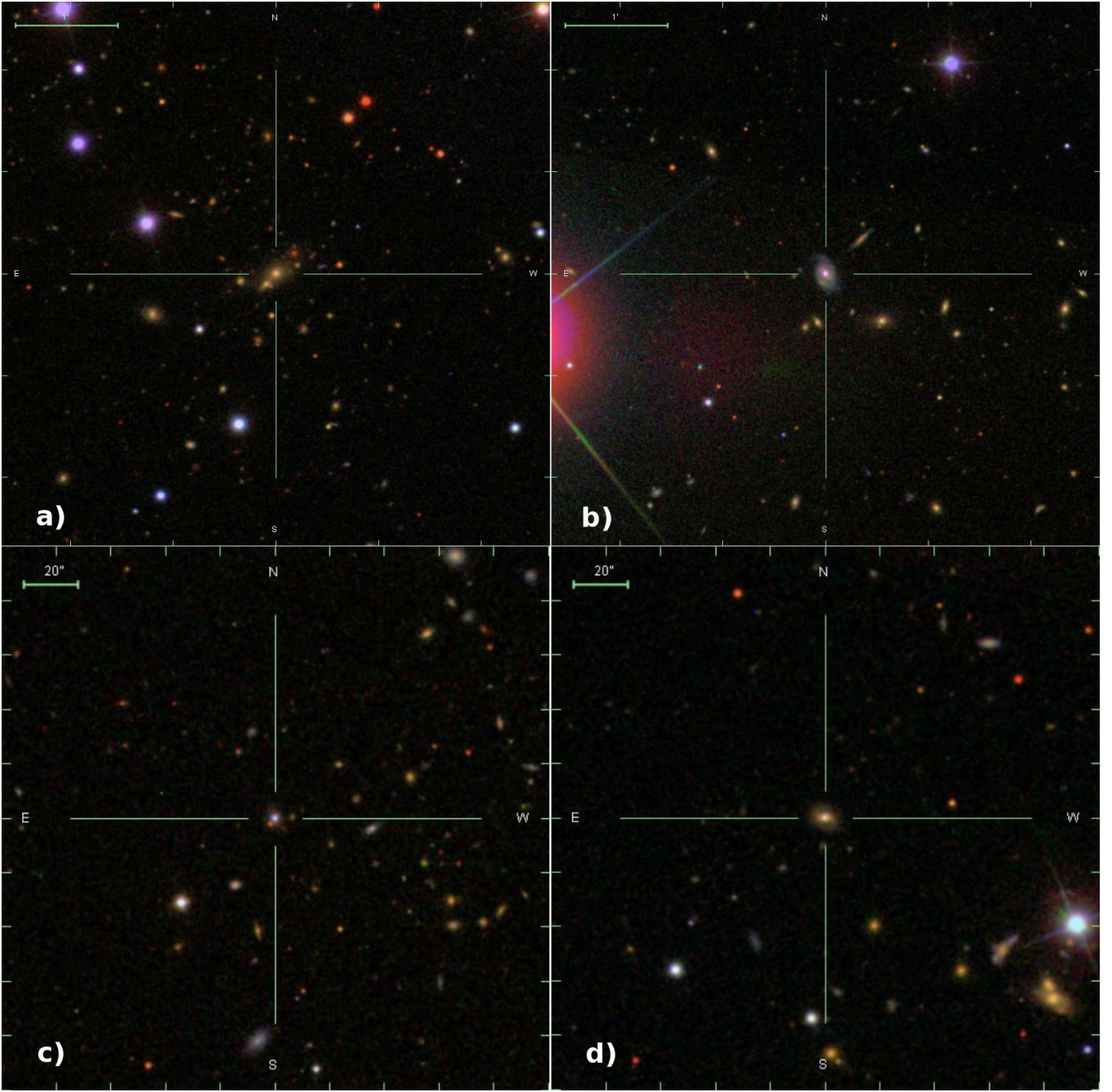

**Figure 8.** Examples of four of the eight different categories used to classify the SDSS g,r,i images of the quasars: a) an early-type like quasar, which is also a BG (E/BG), b) a spiral-like quasar, which is also a BG (Sp/BG), c) an early-type like quasar (E), which is not a BG, d) a spiral-like quasar (Sp), which is not a BG.

A statistical summary of our classification is presented in Table 2. For each of the different radio classes, column 2 gives the total number of quasars (the five unclassified RD quasars are not counted). Column 3 gives the percentage of unresolved, point-like sources (PL) in each sample (relative to the numbers in column 2). On the first line we then give the classification for the unresolved quasars, and put the classification for the resolved one on the line below. For the two main morphological classes, E and Sp (thus, excluding the quasars classified as PL), we give the relative fractions of quasars in each sample that have the considered characteristics (therefore, only the percentages in column 4 and 8 add to 100). For example, of the 1527 RU quasars, 39% are PL, which means that the rest, 61% (924 RU quasars) are resolved. Of these latter, 59% (547) are E type, the rest being Sp. Then, the three other percentages for the E RU quasars are relative to the total 547: 18% are M/I, 21% Is and 15% BG (proceeding similarly for the 41% resolved RU quasars classified as Sp).

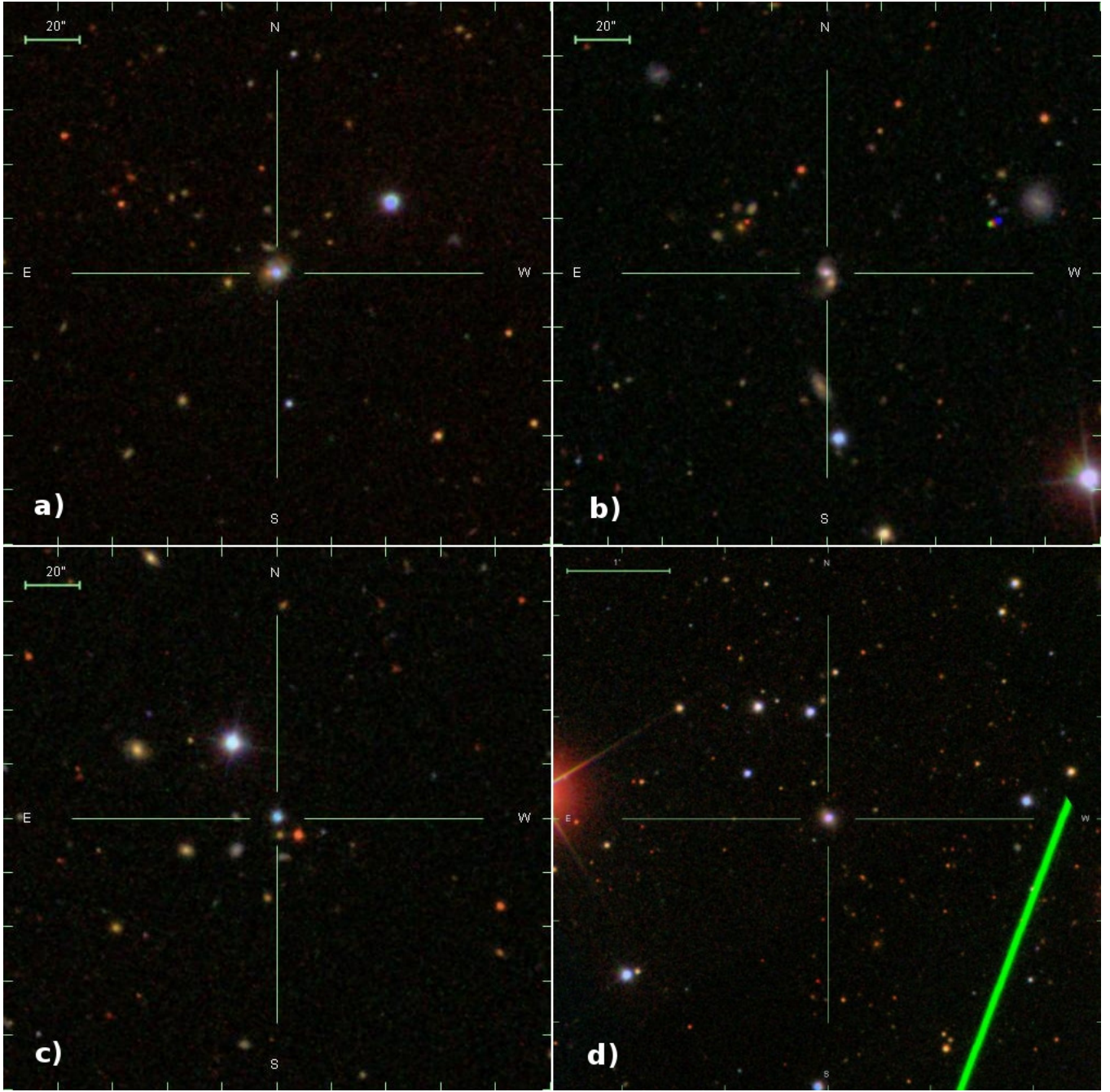

**Figure 9.** Examples of four of the eight different categories used to classify the SDSS g,r,i images of the quasars: a) merging/interacting early-type like quasar (E - M/I), b) merging/interacting spiral-like quasar (Sp - M/I), c) point-like or unresolved quasar (PL), d) apparently isolated quasar (Is).

To quantify the richness of the environments of the quasars in our sample, we also used a special program that was developed to estimate the galaxy density of SDSS galaxies in the Northern Galactic Cap (Ortega-Minakata et al. 2014). Following Wing & Blanton (2011), the galaxy density is parameterised as the number $N_R^{-19}$, which represents the number of galaxies more luminous than absolute magnitude $M_r = -19$, within a projected radius $R$ (in Mpc) around each target object, and within a range in radial velocity of ± 2500 km s$^{-1}$ around the redshift of the target. For the potential neighbour galaxies, the code uses the catalog of galaxies of the SDSS DR7 with available photometric redshifts (http://casjobs.sdss.org), which limits the application of our neighbour search to 90% of the quasars in our sample. To cover different clustering scales, $N_R^{-19}$ was calculated for three different radii, centred on the position of the quasars at their respective redshift: $R = 1.5$, $R = 1$ and $R = 0.5$ Mpc. The corresponding galaxy densities, $N_{0.5}^{-19}$, $N_{1.0}^{-19}$ and $\frac{-19}{1.5}$, in the environment of each quasar in our sample, can be found respectively in columns 10, 11 and 12 of Table 1.

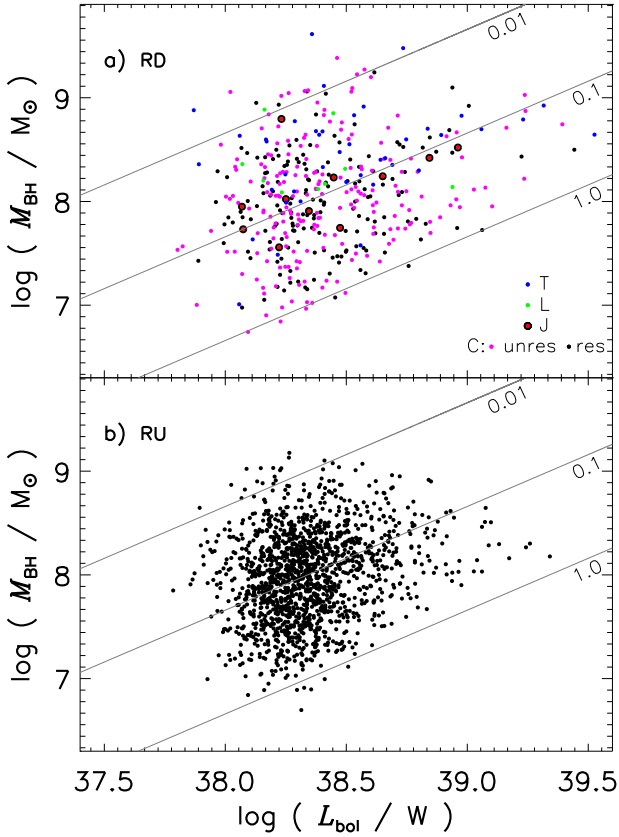

**Figure 10.** The BH mass as a function of the bolometric luminosity for: a) the RD quasars, with different radio structures, and b) the RU quasars. The diagonals correspond to different Eddington ratios $\Gamma = L_{bol}/L_{Edd}$.

## 3 RESULTS

### 3.1 Comparing the characteristics of RU and RD quasars

In Fig. 10 we trace the distribution of the BH mass, $M_{BH}$, as a function of the bolometric luminosity, $L_{bol}$, comparing the RD with the RU quasars. The diagonals correspond to three different Eddington ratios, $\Gamma = L_{bol}/L_{Edd} = 0.01, 0.1$ and 1.0. The BH masses and Eddington ratios of the quasars in our sample are comparable to those of nearby BLAGNs observed in reverberation (Peterson et al. 2004, 2005). No clear distinction appears between the RU and RD quasars, their masses and Eddington ratios being distributed over the same range in both samples.

In Fig. 11a and Fig. 11b we show, respectively, the box and whisker plots for $M_{BH}$ and $\Gamma$, separating the RD quasars according to their radio morphologies. As usual, the bottom and top of the boxes are the first and third quartiles, the bars are the medians and the dots are the means. The whiskers correspond to the lowest/highest datum still within 1.5×IQR of the lower/upper quartile, where IQR is the interquartile range. The notch on the box is a V-shaped region extending to $\pm 1.58 \times IQR/\sqrt{N}$, where $N$ is the size of the sample. Comparing the notches allows one to judge the level of similarity of the medians: overlapping notches suggest the medians have a high probability to be similar. Note that for small samples (e.g., the J quasars) notches can be larger than the boxes.

As it can be seen in Fig. 11a, the RU and C quasars have comparable BH masses. Despite the low number of J and L quasars, we do observe a trend for $M_{BH}$ to increase from the RU and C to the extended radio sources. But only the T quasars have significantly more massive BHs.

In Fig. 11b the RU quasars seem to have lower $\Gamma$ than the C quasars. We also distinguish a statistical trend for $\Gamma$ to decrease from the C quasars to the more extended radio sources, $\Gamma$ being significantly lower in the L and T quasars.

To verify the statistical significance of the differences and better distinguish the trends observed for the means in Fig. 11, we calculated the simultaneous 95% confidence intervals for the pairwise (or family-wise) comparisons of $M_{BH}$ and $\Gamma$ in the different subsamples, using the max-t test (Hothorn, Bretz & Westfall 2008; Herberich, Sikorski & Hothorn 2010). The max-t test is a new parametric test that does not suppose that the variances of the compared samples are the same or that their sizes are similar, which is important in our study. The confidence intervals shown in Fig. 12 are obtained by calculating the difference between two subsample means ($m_{S1} - m_{S2}$), assuming the null hypothesis takes a linear form, adding and subtracting (upper limit and lower limit) the standard error, then multiplying by the value of a t-distribution at a level of confidence of 95%. A confidence interval including zero indicates no statistically significant difference between the subsample means, and the farther from zero the more significant the difference. Applying the max-t test is very simple, as it is part of the many advanced statistical subroutines offered in R[1]. This test is robust and yields results comparable to standard nonparametric tests, like the one-way ANOVA Kruskall-Wallis and Friedman tests, with family-wise post-tests, based on Tukey and Dunnett methods (e.g., Maxwell, Delaney & Kelley 2003; Foster, Barkus & Yavorsky 2006). One interesting advantage of the max-t test is that, since it is based on the comparison of the differences in means, from the signs of these differences one can securily identify the statistical trends, even if the differences are not statistically significant.

In Fig. 12a the max-t test confirms that the T quasars have the most massive BHs, and in Fig. 12b it confirms that they have the lowest $\Gamma$. Although we find no statistically significant differences in BH mass between the C the J and L quasars, the test shows a trend for the mean $\Gamma$ to decrease while the mean $M_{BH}$ increases from the C quasars to the quasars with extended radio structures. Note also the positive sign for the difference C − RU, which is statistically significant. This implies a significant smaller $\Gamma$ on average in the RU than in the C quasars.

Comparing the bolometric luminosity, $L_{bol}$, we observe in Fig. 11c that only the C and T quasars seem to have a higher luminosity than the RU quasars. This is confirmed statistically in Fig. 12c. No statistically significant difference is observed in bolometric luminosity for the RD quasars with different radio morphologies. Based on this comparison, we conclude that RD quasars have higher bolometric luminosity on average than RU quasars.

---

[1] R is a free software environment for statistical computing and graphics: https://www.r-project.org

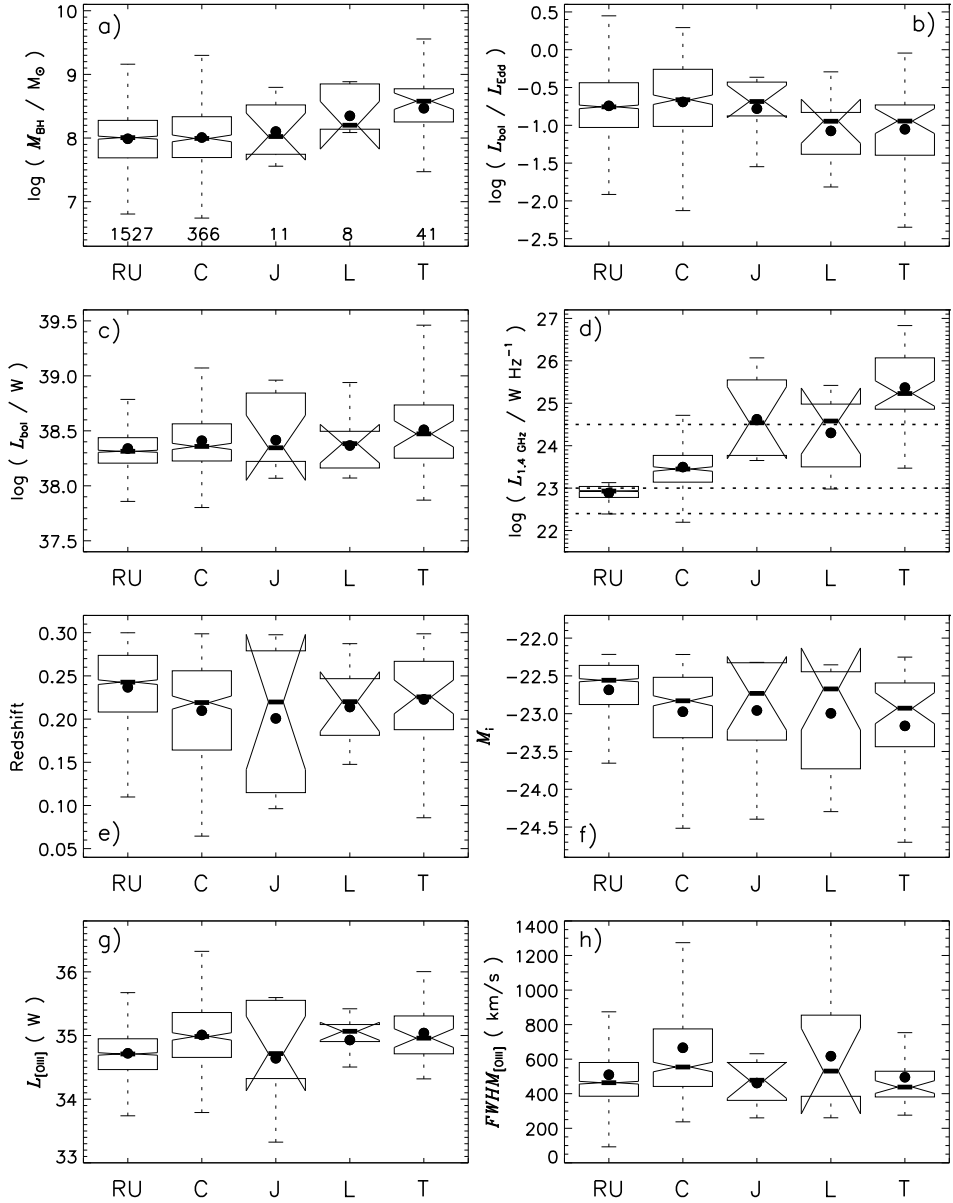

**Figure 11.** Box and whisker plots comparing quasars with different classes of radio structure: a) the masses of the BH, b) the luminosity ratios of quasars, c) the mean bolometric luminosities, d) the radio luminosities at 1.4 GHz; note that for the RU we used the upper limit in flux, while the real measurement from stacking, at the average distance of the RU quasars, is $\mathrm{Log}(L_{1.4\mathrm{GHz}}) \geq 22.4$ W Hz$^{-1}$, e) the distributions in redshift, f) the $i$-band absolute magnitude, $M_i$, g) the luminosities of the [OIII]$\lambda$5007 emission lines, and h) the FWHM of the emission lines [OIII]$\lambda$5007.

Assuming the RU quasars have a radio flux equal to their upper limit, as can be found in col 14 of Table 1, we compare in Fig. 11d and Fig. 12d the radio luminosities of all the quasars at 1.4 GHz, $L_{1.4\mathrm{GHz}}$. Naturally, the C quasars are significantly more luminous in radio than the RU quasars. Note, however, that the stacking yields at the average dis-tance of the RU quasars a luminosity $\log(L_{1.4\mathrm{GHz}}) \geq 22.4$ W Hz$^{-1}$, which is one order of magnitude below the mean for the C quasars, making the difference between the RD and RU quasars even more significant.

What is remarkable in Fig. 11d is the huge difference, by almost two orders of magnitude, in radio luminosity between the C and T quasars. In Fig. 12d, we can also verify that the J and L quasars are significantly more luminous than the C quasars, confirming the differences observed in Fig. 11d. Note that the max-t test finds no statistically significant difference between the J and L quasars, although both tend to be less luminous than the T quasars. Therefore, the trend for $L_{1.4GHz}$ is that it increases along the radio morphological sequence C→J/L→T.

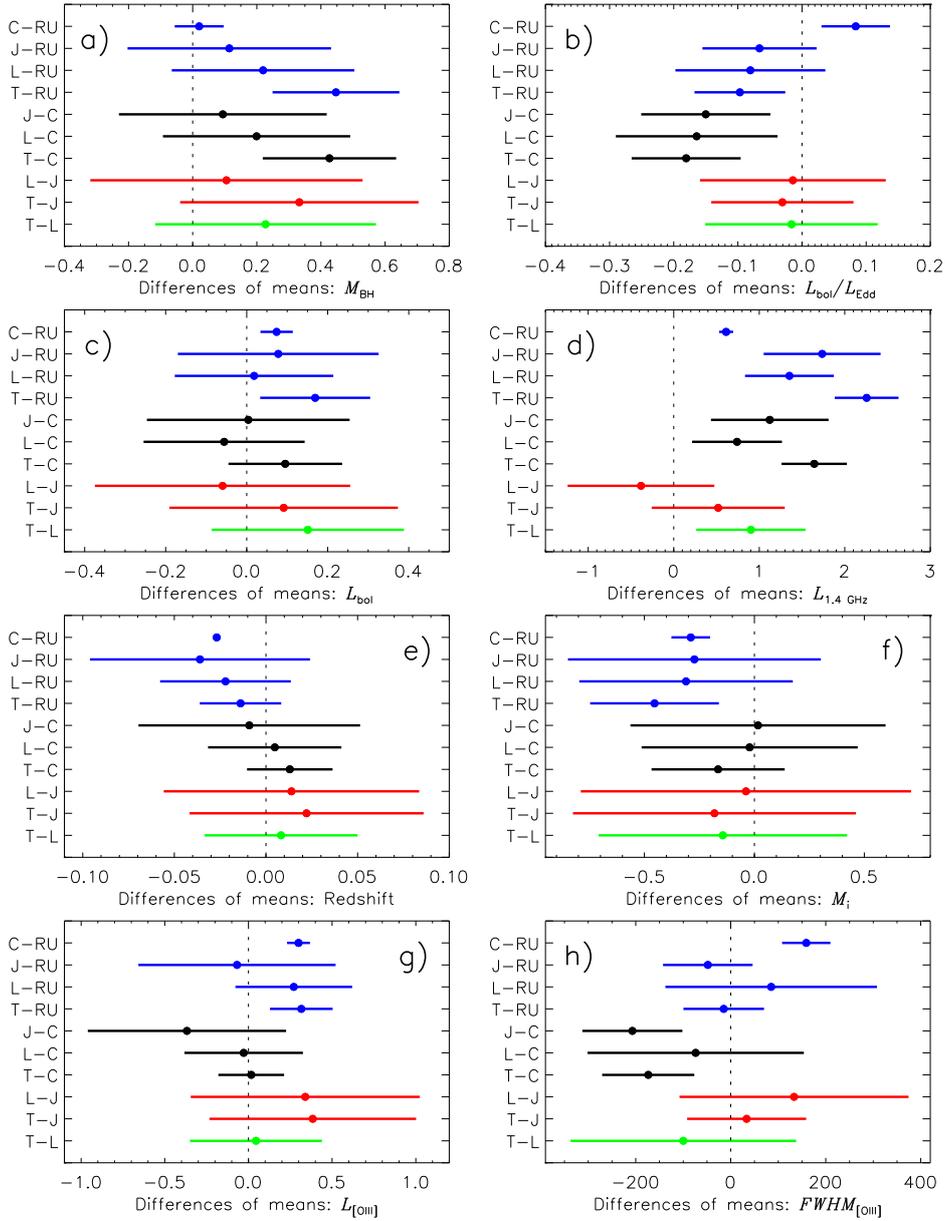

**Figure 12.** Simultaneous 95% confidence intervals for the pairwise comparisons of the subsample means, for the parameters tested in Fig. 11: (blue) relative to RU, (black) relative to C, (red) relative to J, and (green) comparison between T and L. A confidence interval including zero indicates no statistically significant difference between the subsample means, while the signs of the central values yield information on the statistical trends.

Having found some differences between the RU and RD quasars in bolometric and radio luminosities, it is important to verify that these are not the results of a bias due to the distances. Indeed, the distribution in redshift of the quasars in our sample is not homogeneous, the majority of the quasars being located between $z = 0.2$ and $z = 0.3$; the median is $z = 0.27$, near the upper redshift limit of our sample. This implies that the statistics at low redshift may be prone to high fluctuations, due to the small number of quasars, and because quasars emitting in radio are easier to detect at low redshift.

Examining Fig. 11e we do observe a trend for the RD quasars to be found at lower redshift than the RU quasars. However, the max-t test in Fig. 12e only finds a significant difference between the RU and the C quasars, and no significant difference is observed for the RD quasars with different radio morphologies. Based on this result we conclude that the differences in luminosities observed between the various types of quasars in our sample are most probably genuine, reflecting intrinsic physical differences.

Consistent with the above conclusion, the differences in

*i*-band absolute magnitudes, $M_i$, observed in Fig. 11f, suggest that the RD quasars are also more luminous in optical than the RU quasars. This is confirmed statistically in Fig. 12f for the C and T quasars, while, due to the small-number statistics, the difference only appears as a trend for the J and L quasars. No statistically significant difference is found for the RD quasars with different radio morphologies, although there is a weak trend for the T quasars to be more luminous than the C quasars.

Two parameters, the luminosity of the emission line [OIII]$\lambda5007$, $L_{\text{[OIII]}}$, and the FWHM of this line, $FWHM_{\text{[OIII]}}$, define the emission-line characteristics of the quasars. However, the ionised gas producing the [OIII]$\lambda5007$ emission lines is known to extend over much larger regions than the broad lines. Indeed, the average $FWHM_{\text{[OIII]}}$ of the quasars in our sample is 600 km s$^{-1}$, which is typical of extended narrow-line regions in AGNs. In Fig. 11g we observe that, except for the J quasars, the general trend is for RD quasars to have a higher $L_{\text{[OIII]}}$ than RU quasars. This is confirmed statistically in Fig. 12g for the C and T quasars, but only appears as a trend for the L quasars. From this, we conclude that the C and T quasars produce higher fluxes of ionising photons than the RU quasars. No statistically significant difference is observed between the RD quasars with different radio morphologies, in particular between the T and C quasars.

Interestingly in Fig. 11h we find that only the C quasars have larger $FWHM_{\text{[OIII]}}$ than the RU quasars. This difference is confirmed statistically in Fig. 12h. In these two figures, we can also see that the C quasars have higher $FWHM_{\text{[OIII]}}$ than in the other RD quasars (although this only appears as a trend for the L quasars). No statistically significant difference is observed between the other types of RD quasars.

## 3.2 Coupling between the radio morphologies and optical spectral characteristics of the quasars

The estimation of the mass of the BH in our study depends on two parameters related to the emission-line characteristics of the quasars, the luminosity of the continuum at 5100 Å (taken as a proxy for the radius of the BLR) and the FWHM of the broad component of the H$\beta$ Balmer line (taken as a proxy for the gas velocity). To determine which of these parameters explain the difference in BH mass observed in Fig. 11a and Fig. 12a, we have applied to our data a one way ANOVA test with Tukey's post tests (Maxwell, Delaney & Kelley 2003; Foster, Barkus & Yavorsky 2006). The ANOVA test confirms a significant difference of $FWHM_{\text{H}\beta}$, the post-tests showing that the T quasars have higher $FWHM_{\text{H}\beta}$ than both the RU and C quasars, which have comparable $FWHM_{\text{H}\beta}$.

Similar results where obtained before by Zamfir, Sulentic & Marziani (2008), comparing quasars above and below the radio luminosity boundary log $L_{1.4\text{GHz}} \sim 24.5$ W Hz$^{-1}$. These authors showed that quasars with high radio luminosities have higher $FWHM_{\text{H}\beta}$ than quasars with low radio luminosities. In their analysis, these authors also claimed that quasars with different radio power show different values of the emission-line index, $R_{\text{Fe II}}$, the powerful radio sources being limited to a narrower range of low $R_{\text{Fe II}}$ indices than the weak sources.

In our analysis, the relation found by Zamfir, Sulentic & Marziani (2008) might suggest a possible connection between the radio morphology and the optical spectral characteristics of the quasars. In Fig. 11d, we traced the radio luminosity boundary proposed by Zamfir, Sulentic & Marziani (2008). One can see that such definition does separate the T from the C quasars. Interestingly, we also observe, based on Fig. 11h and confirmed in Fig. 12h, a systematic difference in the optical spectra of the C and T quasars, the C type having higher $FWHM_{\text{[OIII]}}$ than the T type.

Could our observations be consistent with those of Zamfir, Sulentic & Marziani (2008)? Fortunately, although these authors give no explanation for the behavior of $R_{\text{Fe II}}$, we, on the other hand, can propose one for the behavior of $FWHM_{\text{[OIII]}}$. Indeed, it is reported in Osterbrock & Ferland (2006) that the FWHM of emission lines in narrow-line AGNs tend to increase as the ionisation potential increases (see their figure 14.10). In other words, the higher the level of ionisation, the higher the FWHM. This, the authors explained, can be due to the hardening of the ionising photon continuum as we approach the ionising source (the BH in an AGN). Consistent with this view, Osterbrock & Ferland (2006) also showed that emission lines with the same level of ionisation will also tend to have a higher FWHM if the regions where they form have higher density. These two effects combined would thus explain why the $FWHM_{\text{[OIII]}}$ of the C quasars is higher than for the T quasars. This is because in the C quasars the narrow-line regions are possibly more concentrated than in the T quasars, and thus nearer to the BH where the flux of ionising photons is harder and the density higher. This result does suggest that the ionising fluxes produced by the quasars are coupled to the radio structures.

But is this the same effect observed by Zamfir, Sulentic & Marziani (2008)? Explaining the behavior of the $R_{\text{Fe II}}$ is more difficult, because this index is complex. By definition, $R_{\text{Fe II}}$ is equal to the ratio of the equivalent width of the iron Fe II line to the equivalent width of the H$\beta$ line, EW(Fe II)/EW(H$\beta$), and, thus, it depends on four different parameters, the two fluxes for the emission lines and the two adjacent fluxes for the continuum. However, it is easy to realize that the simplest way to increase $R_{\text{Fe II}}$ would be to increase the intensity of the Fe II line relative to its continuum. This, on the other hand, implies a hardening of the ionising photon spectrum and/or increase in gas density. This is exactly what we expect for compact ionised regions near the center of the AGN (Osterbrock & Ferland 2006). Therefore, it is highly probable that our observations indeed concur with those of Zamfir, Sulentic & Marziani (2008).

If our interpretation is correct, then, and consistent with the claim made by Zamfir, Sulentic & Marziani (2008), the optical spectral characteristics of the quasars are, indeed, coupled to their different radio morphologies.

## 3.3 Comparing environments and galaxy host morphologies

In Fig. 13 we show the box and whisker plots for $N_{\text{R}}^{-19}$ as obtained using the three different projected search radii. Comparing the galaxy densities in the environments of the RD and the RU quasars with different radio morphologies we distinguish no difference. This is true for any value of the search radius. The absence of variations in galaxy densities for quasars with different radio morphologies is confirmed statistically by the results of the max-t test in Fig. 14. Median values of $N_{1.5\mathrm{Mpc}}^{-19}$ of the order of 8 to 10 suggest the quasars are forming part of loose groups of galaxies, which is a common field structure at low redshift (cf. Ortega-Minakata et al. 2016, in preparation).

The results of our eye examination to determine the host galaxy morphology of each of the 1953 quasars in our sample (the five unclassified RD quasars identified as X in Table 1 are not counted) can be found in Table 2. In general, our results are consistent with those obtained by Falomo et al. (2014), who applied a rigorous photometric study on deep images of 400 nearby quasars in the SDSS Stripe 82. For example, they found as a whole 58% bulge-dominated quasar hosts (37% bulge-dominated quasars and 21% with a bulge and disc component), which compares very well to our 59/52% early-type (E) RU/C quasars. Additionally, we classified 41/48% RU/C quasars as spiral-like (Sp), compared to 42% according to Falomo et al. (2014), which, again, is remarkably similar. Interestingly, our fraction of possible merging/interacting RU/C cases, ∼ 18/35% for the E and 17/24% for the Sp, is also comparable to their fraction of quasars with a complex structure, which in total is about 32%.

Comparing the RD quasars with the RU quasars we deduce from our eye inspection that there is no significant difference in the fraction of optically unresolved (PL) quasars (the only exception are the J quasars, which show only one PL). Comparing the resolved C and RU quasars, we find no significant difference in the fractions of E and Sp quasars. On the other hand, all the J and L quasars are of type E, but those are relatively small samples. There is also a trend for the T quasars to have a higher number of E than the RU and C quasars, but, again, this trend is statistically marginal.

The fraction of isolated quasars (Is) is also comparable in all the samples, which is consistent with the previous conclusion based on $N_{\mathrm{R}}^{-19}$. All the quasars in our sample seem to reside in similar environments, consistent with low density galaxy structures, like loose groups.

Another interesting difference noted in Table 2 is that there seems to be a slightly higher fraction of possible merging or interacting cases (M/I) in the C, J and L quasars than in the RU quasars sample. However, this is not observed for the T quasars.

The results in Table 2 may also suggest that RD quasars have a significantly higher probability than RU quasars to be among the brightest galaxies of the structure they are found in (identified as BGs in Table 2). However, except for the T quasars, the fraction of BGs is well below 50%, which suggests that, in general, there is no obvious difference between the RD and RU quasars, both are hosted in similar types of galaxies. This seems to confirm our conclusion of low density galaxy structures based on $N_{\mathrm{R}}^{-19}$.

In general, therefore, we conclude that there are no obvious differences between the RU and RD, or between the RD with different radio morphologies, of galaxy host morphology and number of merging galaxies.

The above results suggest that the difference in *i*-band absolute magnitude, as observed in Fig. 11f and Fig. 12f, is not related to the galaxy hosts, but, rather, to the activity of the BH itself. Indeed, in studies of galaxies, the *i*-band absolute magnitude is usually taken as an indicator of the stellar mass. Interpreted in this way, the differences between the RU and RD quasars would suggest that the former reside in less massive galaxies than the latter. But we have no evidence for that. For example, assuming the BH formation is linked to the formation of the galaxies, one would thus expect the RU quasars to have smaller mass BHs than the RD quasars, which is not observed in Fig. 11a and Fig. 12a. Only the T quasars have significantly higher mass BHs than the RU quasars, and no significant difference in *i*-band absolute magnitude is detected for the RD quasars with different radio morphologies. On the other hand, the RD quasars show higher bolometric luminosity and ionizing fluxes than the RU quasars, which suggests that the higher *i*-band absolute magnitude of the RD quasars may also be due to their higher level of AGN activity.

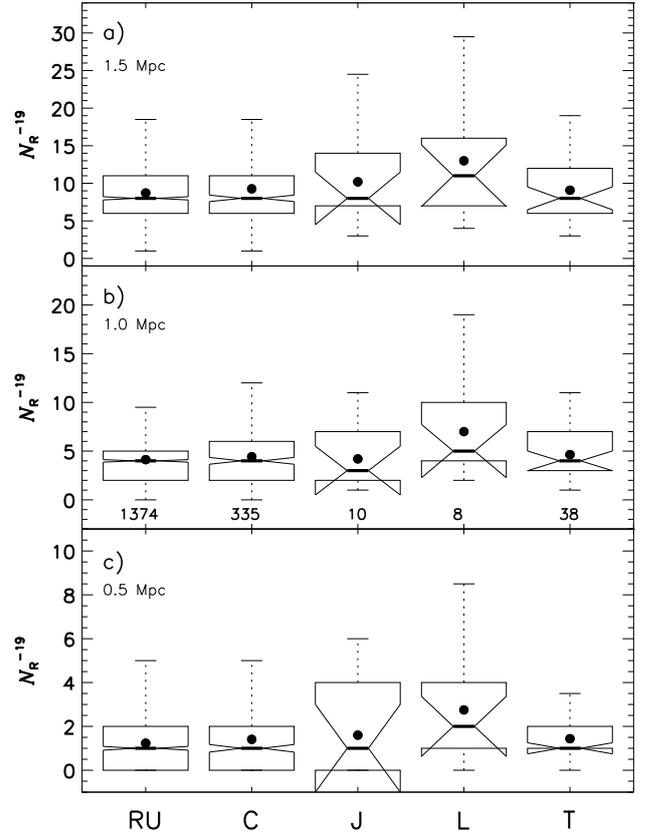

**Figure 13.** Box and whisker plots comparing the galaxy density indices $N_{\mathrm{R}}^{-19}$ for quasars with different classes of radio structure in the SDSS Northern Galactic Cap (90% of our sample); a) $R = 1.5$ Mpc, b) $R = 1$ Mpc and c) $R = 0.5$ Mpc.

## 4 DISCUSSION

### 4.1 Definitions of Radio Loud and Powerful Radio Sources

Since the goal of our discussion is to search for new clues to explain the difference between radio-loud (RL) and radio-quiet (RQ) quasars, we must first establish how we distinguish between the two. This exercise is not trivial, since many different definitions can be found in the literature, which can introduce some confusion during our discussion.

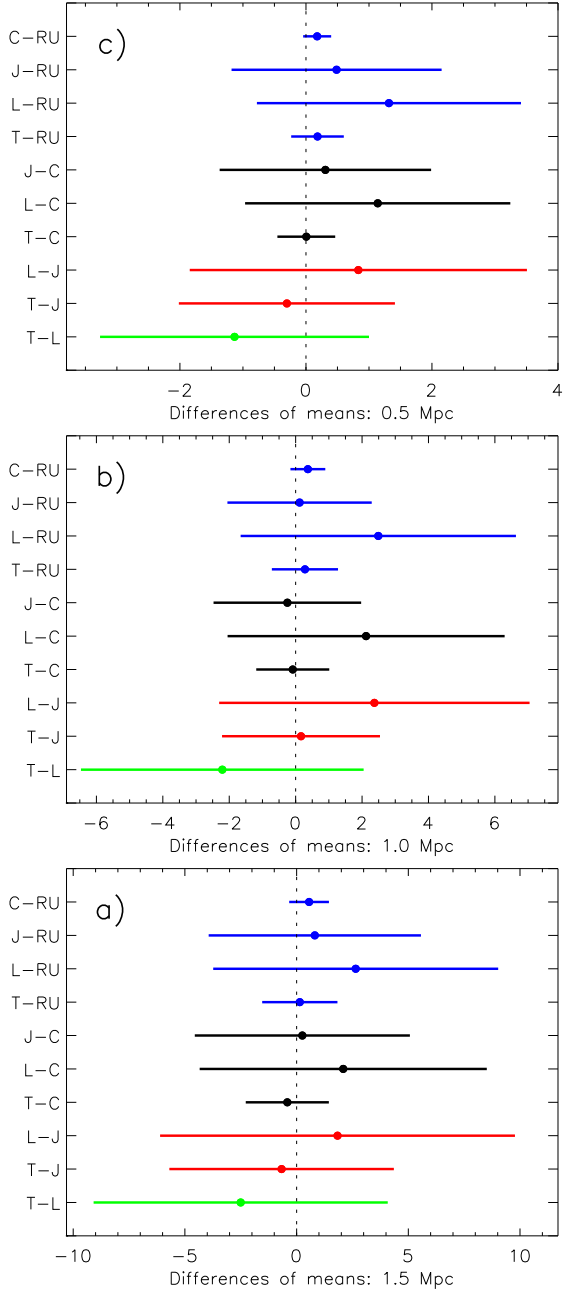

**Figure 14.** Simultaneous 95% confidence intervals for the pairwise comparisons of the mean in galaxy density indices $N_R^{-19}$ for a) $R = 1.5$ Mpc, b) $R = 1$ Mpc and c) $R = 0.5$ Mpc.

For example, in their study of SDSS radio galaxies (excluding quasars), Best et al. (2005a) define RL galaxies as those galaxies that have a radio luminosity $\log(L_{1.4GHz}) \geq 23$ W Hz$^{-1}$. They legitimate this definition by stating that this luminosity marks a natural boundary above which the AGN becomes the principal source of radio emission (taking into account the possible contribution of star formation; see also Best et al. 2005b). In Fig. 11d, we traced this luminosity criterion over the box and whisker plots comparing the radio luminosities of the quasars in our sample. Most of the quasars classified as RD, 85%, are above this luminosity, while 98% have $\log(L_{1.4GHz}) \geq 22.5$ W Hz$^{-1}$. Therefore, adopting this last, slightly lower value, we can confidently classify all the RD quasars in our sample as RL, and all the RU as RQ. Based on this definition, we conclude that about 80% of the nearby quasars are RQ, which is consistent with what we know about quasars in the Universe.

On the other hand, our stacking result also suggests that RQ quasars are not "radio-dead", only that they emit in radio at a level too faint to be detected in current large-scale radio surveys. Obviously, this implies that by increasing the sensitivity of future surveys, many more radio sources in quasars could be found. However, this would not change the definition of RL and RQ galaxies, since most of these new radio sources would have radio luminosities below $\log(L_{1.4GHz}) < 22.5$ W Hz$^{-1}$ (see, for instance, Kellermann et al. 2016).

Another important distinction is frequently made in the literature between powerful (PRS) and weak (WRS) radio sources (e.g., Blandford et al. 1990). Usually the distinction is made by applying a radio luminosity limit, for example, $\log(L_{1.4GHz}) = 24.5$ W Hz$^{-1}$ (Blandford et al. 1990). Tracing this luminosity criterion in Fig. 11d suggests that most of the RL quasars that have an extended radio structure in our sample, namely, all the T quasars and almost half of the J and L quasars, are PRS. Adopting this definition, then about 3% of the nearby quasars in our sample would be PRS, which, again, is consistent with what we know about quasars in the Universe.

We conclude, therefore, that by adopting simple luminosity criteria, our sample of quasars can be separated in RQ/RL and WRS/PRS in a way which is consistent with what is generally known about quasars in the literature.

Alternatively, many authors have used $R$, the ratio of radio-to-optical (originally B band) luminosity to distinguish between RL and RG AGNs. Based on this parameter, it is proposed that RL AGNs must have $\log(R) \geq 1.0$ (e.g., Kellermann et al. 1989), and some even suggested $\log(R) \geq 2.0$ (e.g., Liu, Jiang & Gu 2006; Zamfir, Sulentic & Marziani 2008). In Fig. 15a, we trace the radio luminosity as a function of the $i$-band luminosity (see Ivezić et al. 2002, for a similar figure), as observed for the RL quasars in our sample (classified based on their radio morphologies). We find no segregation of data on opposite sides of the line $\log(R_i) = 1$. In Fig. 15b we trace the same diagram for the RQ quasars in our sample, using their upper limits as radio fluxes. Comparing with Fig. 15a, we find no evidence of a separation between the RD and RQ quasars based on the criterion $\log(R_i) \geq 1.0$. Considering that the mean luminosity for the stacking result is $\log(L_{1.4GHz}) = 22.4$ W Hz$^{-1}$, the impression given by the distributions of $R_i$ is one of a continuous distribution (for similar results see Cirasuolo et al. 2003; Rafter et al. 2009).

However, we do observe in Fig. 15a that most (not all) of the RL quasars with a C radio structure are below $\log(R_i) = 2.0$, while almost all those that have an extended structure, J, L and T are above this value (see also Rafter et al. 2011). Note that a boundary value of $\log(R_i) = 2.0$ is roughly consistent with a boundary in luminosity $\log(L_{1.4GHz}) \sim 24.5$ W Hz$^{-1}$ which is the luminosity limit we adopted to separate WRS from PRS.

Although it is clear based on Fig. 15a that the criterion

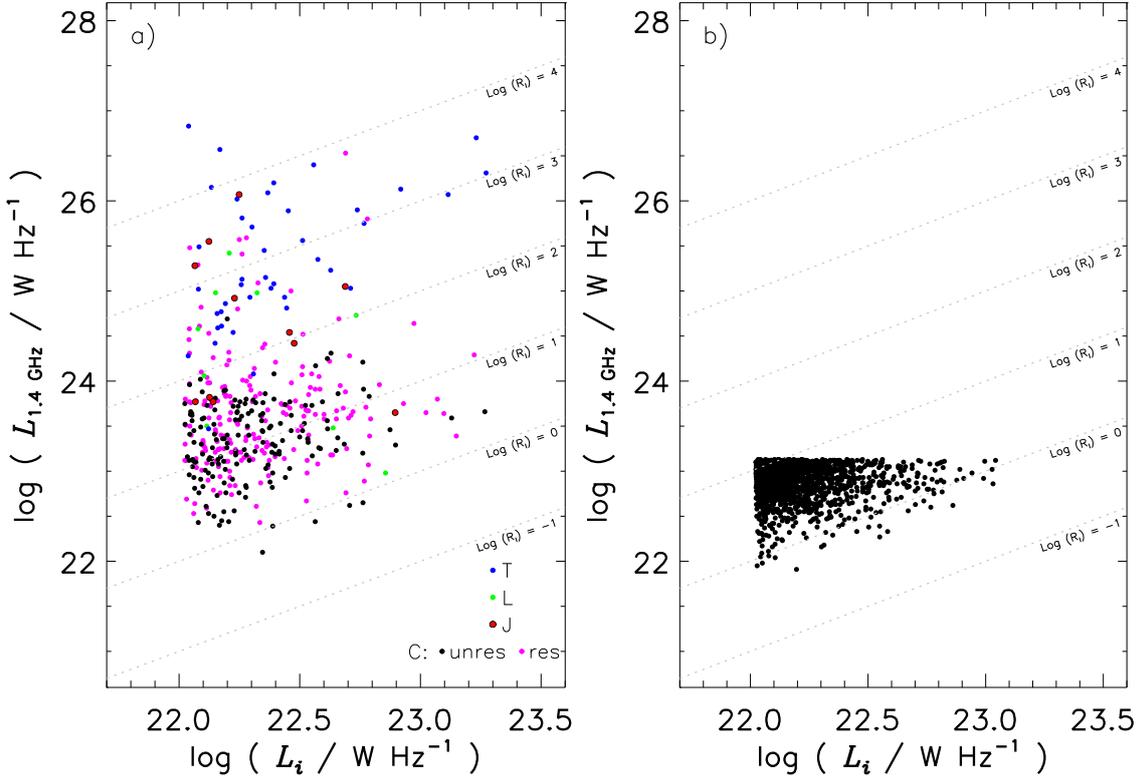

**Figure 15.** Radio luminosity at 1.4 GHz as a function of luminosity in the *i*-band for a) the RL quasars in our sample, b) for the RQ quasars we used the upper limits in flux as found in col 14 of Table 1. The diagonal lines correspond to different ratios (in log) of radio luminosity to *i*-band luminosity, $R_i$.

$\log(R_i) = 2.0$ does not allow an unambiguous separation between the WRS and PRS quasars in our sample, we must admit that the trend is definitely there. As we have commented in Section 3.2, Zamfir, Sulentic & Marziani (2008) went farther by showing that the WRS and PRS (but using the terms RQ and RL instead of WRS and PRS) have also different $R_{Fe\,II}$ index. In Section 3.2, we have recognized a similar difference in the optical spectra of the WRS and PRS, in terms of the $FWHM_{[OIII]}$, and interpreted it as evidence for a coupling between the optical spectral characteristics of the quasars and their radio morphologies. In particular, the apparent bimodalility of the LLS distribution, in Section 2.2 (Fig. 6), which seems to separate, at least in part, the compact WRS from the extended PRS, would also be consistent with the separation of these objects based on the $\log(R_i) = 2.0$ criterion. Therefore, the distinction between WRS and PRS, like emphasized by Zamfir, Sulentic & Marziani (2008), seems an important distinction to make in order to better understand what sparks the radio-loud phase of quasars.

Note that another kind of bimodality was proposed recently by Kellermann et al. (2016). Based on a deep radio survey with the VLA at 6 GHz of 178 quasars with redshifts between $0.2 \leq z \leq 0.3$, they postulated that RQ and RL quasars form two quasars populations due to two different sources of radio emission: above $\log(L_{6GHz}) \sim 23$ W Hz$^{-1}$, BHs are the primary sources of radio emission in RL quasars, while below this luminosity, star formation in the RQ quasar galaxy hosts are the dominant radio sources. Note that such interpretation looks similar to the one proposed by Best et al. (2005a) to separate RL from RQ galaxies. However, having adopted a separation criterion similar to these last authors to separate the RL and RQ quasars in our sample, our stacking analysis and the comparison of the $R_i$ criterion in Fig. 15 show no evidence for two quasars populations.

To test the star formation hypothesis further, we can use data in mid infrared (MIR) and search for differences in colors. Indeed, as was shown in Coziol, Torres-Papaqui & Andernach (2015), it is possible to distinguish the level of star formation in quasars using a new diagnostic diagram based on WISE colors. Having in hand the MIR colors of the 1958 quasars in our sample, we have cross-correlated our list to the list of Kellermann et al. (2016) and found all of their quasars, except seven. In Fig. 16a, we compare the MIR colors of the quasars classified as RL and RQ according to Kellermann et al. (2016) luminosity criterion. We found no evidence for two quasar populations separated by the level of star formation in their host galaxies. In fact, the bulk of these quasars are located in what is defined as the low star formation zone (low SF zone; see Coziol, Torres-Papaqui & Andernach 2015, for how the different zones were determined in the diagnostic diagram).

In Fig. 16b, c and d we now compare the wise colors of all the RQ and RL quasars classified by us based on their different radio morphologies. There is a clear trend for the quasars with extended radio sources to be located in the low SF zone, while the compact radio sources tend to be located in the high SF zone. However, the RQ show no such prefer-ences. Interestingly, we verified that the positions occupied by the different quasars in this diagnostic diagram are not correlated with the galaxy host morphologies, which is consistent with the analysis presented in Section 3.3. Therefore, although we do distinguish a potentially higher level of star formation in the WRS than in the PRS, we must conclude that, in general, there is no clear evidence of different levels of star formation in the RQ and RL quasars taken as a whole.

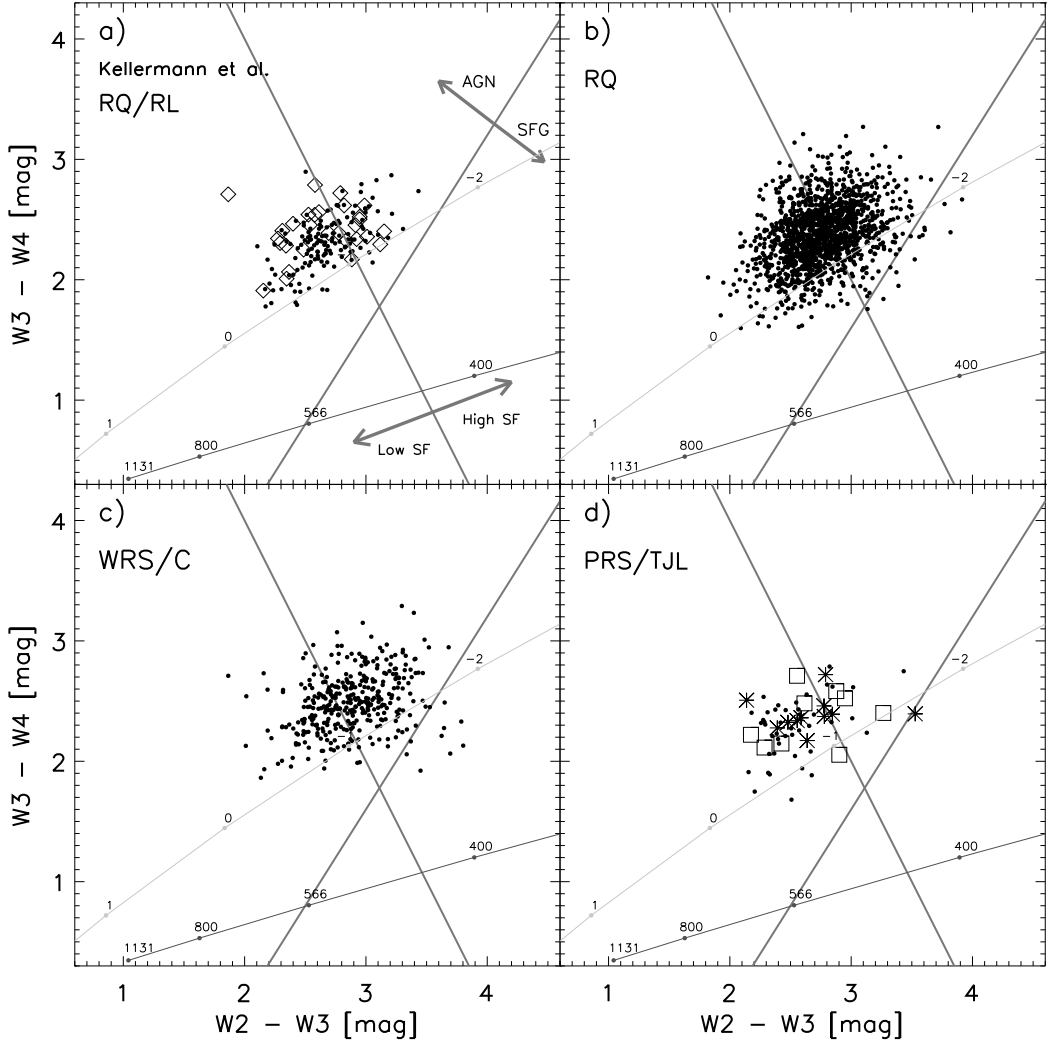

**Figure 16.** Diagnostic diagram using WISE colors as proposed by Coziol, Torres-Papaqui & Andernach (2015): a) the sample of Kellermann et al. (2016) separated in RL (diamonds) and RQ (black dots) using their own luminosity criterion, b), c) and d, all the quasars in our sample separated based on their radio morphologies; b) RQ, c) C type, which are all WRS, and d) extended radio sources, J (square), L (stars) and T (black dots), which are almost all PRS.

### 4.2   What sparks the radio-loud phase of nearby quasars?

Now that we have clarified our definition of RL quasars and emphasized the importance of distinguishing between WRS and PRS, we are ready to discuss what the results of our analysis can tell us about the phenomenon that sparks the radio loud phase in nearby quasars?

The different hypotheses described in our introduction can be separated in two types: 1) the "accidental difference" conjecture, and 2) the intrinsic difference conjecture. The first of the accidental difference conjecture proposes that RL quasars are observed near face on, making it a core dominated radio source, or at intermediate angles relative to our line of sight, where the radio core and lobes are all visible (e.g. Barthel 1989). This hypothesis is very much similar to the unification model used to explain the differences between narrow and broad-line AGNs, as observed in the optical (Antonucci 1993; Urry & Padovani 1995). However, our selection criterion, based on the optical properties of the quasars, deliberately favour the face on orientation for the RQ and RL quasars. Despite this selection bias, we observe the same fractions of RL to RQ and of PRS to WRS as observed generally for quasars. Therefore, either we must assume the orientation angle in radio differs drastically from that in the optical, which go against the original proposition, or we must conclude that the orientation angle does not play an important role in radio. In favour of the latter possibility, our analysis suggests that the RL and RQ quasars form a continuous distribution, differing only in their levels of radio emission, whereas considering the differences between the WRS and PRS, a difference in BH masses clearly suggests that the difference must be intrinsic to the radio sources (for a similar conclusion, see discussion in Kellermann et al. 2016).

Could evidence of boosting clarify the orientation angle problem (Kellermann et al. 1989; Cirasuolo et al. 2003; Lu et al. 2007)? This is difficult to verify with our data, because our selection criterion eliminates the most extreme cases, namely, the blazars. However, if boosting in radio is nonetheless present, we would expect it to appear in the most compact radio sources, more specifically, those quasars with a C type that are unresolved in radio. In Fig. 15a, we have separated the C subsample in resolved and unresolved radio sources. Doppler boosting in such objects would be expected to produce higher $R_i$ values. Contrary to our expectation, we see that the resolved and unresolved C quasars have similar $R_i$, and all those that have an untypically high $R_i$ are resolved, which is consistent with the possibility of undetected extended structures in these sources (see Section 2.2). Therefore, there is no clear evidence of Doppler boosting in our sample.

The second of the accidental difference conjecture proposes that the radio efficiency is transient, which implies that RQ quasars would eventually evolve into RL and, assuming the active radio phase is of short duration, would then transformed back into RQ sources (e.g., Broderick & Fender 2011; Saripalli et al. 2012). However, all the differences we observe between the RQ and RL quasars, more specifically, the higher bolometric luminosities and ionizing fluxes in RL, and the higher BH masses for the T type, would then imply that most RQ quasars in our sample are in a pre-active radio phase. Such "bias" is not consistent with our selection criterion.

Consequently, the intrinsic difference conjecture, namely, that RQ and RL quasars must have different central engines, seems like a better hypothesis. However, here too we can distinguish between two general trends, the first suggesting that the BHs itself must differ, more specifically, being more massive or spinning more rapidly in RL than in RQ quasars, while the second is suggesting that intricate details of the accretion process explain the higher radio emission efficiency of RL quasars.

One new clue concerning the intrinsic difference conjecture comes from the comparison of the masses of the BHs. In Section 3.1, we have found that the RQ and RL quasars, the latter dominated by the C type, show no difference in BH masses. Only the extended T type have more massive BHs. Note, however, that this difference is small, compared to the difference in radio luminosity, the BHs in the PRS being about three times more massive on average than the BHs in the RQ and quasars, while, at the same time, the radio luminosities of the PRS is found to differ by almost two orders of magnitude from the radio luminosities of the WRS. Therefore, although there is a clear evidence for a difference of BH masses, it is not obvious how this difference is related to the extended radio structures. Moreover, the fact that we observe no difference in terms of host galaxy morphologies, levels of interaction, and environments, strongly suggests that what distinguishes the RQ from the RL quasars cannot be related to a generic process, like any processes related to the formation of galaxies.

Because there is no direct observation related to the spin of a BH, the spin paradigm, namely that BHs rotate faster in RL than in RQ quasars, cannot be tested directly in our analysis. However, theoretically, rotating BHs are the most general solution of Einstein's equations, therefore most BHs rotate. Moreover, Blandford & Znajek (1977) have demonstrated that an enormous amount of energy can be extracted from a rotating BH by magnetic braking. Therefore, in the majority of quasars an important part of the energy must come from the spin of their BHs. Now, based on the definition of the irreducible mass (Frolov & Zelnikov 2012) it can be shown that the higher the angular momentum of the BH, the higher the amount of energy that can be extracted from it. However, it is not clear how this reducible process transforms into more efficient radio jets. Another problem is why, based on such a general phenomenon as the spin of BH, so few quasars are RL? Assuming the spin paradigm is correct, our analysis would imply that most of the BHs in the RQ quasars are slow rotators. What threshold in spin would explain that? Then ~20% are RL, but only ~3% have powerful radio jets. What spin value would produce the difference between the WRS and PRS? More importantly, what mechanism would explain the difference of spins? Most models in the literature assume generic processes related to the formation of galaxies (e.g., Fanidakis et al. 2011). But such processes should also be general, by definition, and, consequently, it might be difficult to reconcile these processes with the fact that there are very few RL and PRS quasars. On this point our analysis of the morphologies and environments of the different quasars in our sample shows no evidence suggesting a difference in their formation processes.

In King & Pringle (2006) the author explained that for super-massive BHs to form over relatively short interval of time, as suggested by observations at high redshifts, low radiative accretion efficiencies, and thus low BH spins, are preferred. This is easy to understand. If we assume the general formula for the bolometric luminosity due to accretion of matter onto a BH is (Frank, King & Raine 1992):

$$L_{\rm bol} = \eta \dot{M} c^2 \tag{11}$$

where $\dot{M}$ is the mass accretion rate (in units of $\rm M_\odot/yr$) and $\eta$ is the radiative efficiency of the accretion process (with a standard value of 0.1, but, in reality, an uncertain parameter, since it depends on details of the disc accretion model, e.g., Blandford & Znajek 1977; Frank, King & Raine 1992), then its mass must grow as (King & Pringle 2006):

$$\dot{M}_{\rm BH} = (1 - \eta)\dot{M} \tag{12}$$

Thus, the higher the radiative efficiency, the lower the mass accretion rate on the BH. Note that King & Pringle's argument applies perfectly to the RQ quasars, which are supposed to be slow rotators. However, this would not explain the difference between the RQ and RL quasars (since we see no difference of BH mass between the RQ and C type RL quasars), unless, maybe, the mass accretion process is unrelated to the process producing the higher radio efficiency. But this contradicts our evidence for a coupling between the two phenomena, as discussed in Section 3.2, and may also pose a problem considering that only the PRS quasars show, at the same time as their higher radio luminosities, significantly more massive BHs than the WRS quasars.

Which new clues can our analysis provides about this problem? We have found that, in general, RL quasars produce more energy than RQ quasars (Ghisellini 1993; della Ceca et al. 1994; Ciliegi et al. 1995; Daly 1995; Wu et al. 2002; Bian & Zhao 2003; Celotti 2005; Balmaverde et al. 2008; Koziel-Wierzbowska & Stasińska 2011). This is not only obvious in radio, but also in the optical based on $L_{bol}$ and $M_i$, and even based on $L_{[OIII]}$, which is a parameter related to the number of ionizing photons. In terms of the BH activity, these characteristics, according to equation (11), must imply higher accretion rates or radiative efficiencies (or both) in the RL than in the RQ quasars.

Evidence that this is indeed the case can be found in the different Eddington ratios of quasars with different radio morphologies. Since the Eddington luminosity depends on the BH mass, $L_{Edd} \simeq 10^{31} (M_{BH}/M_\odot)$ W, we obtain for the Eddington ratio the following relation:

$$\Gamma = \frac{L_{bol}}{L_{Edd}} = 10^{-31} \left(\frac{\eta \dot{M}}{M_{BH}}\right) c^2 \sim 5.71 \times 10^8 \left(\frac{\eta \dot{M}}{M_{BH}}\right) \quad (13)$$

Therefore, since we found the RQ quasars to have significantly lower $\Gamma$ than the C type quasars, but comparable BH masses (cf. Fig. 12b and Fig. 12a), then according to equation (13) the C type quasars must have higher accretion rates and/or radiative efficiencies than the RQ quasars, that is, $(\eta \times \dot{M})_C > (\eta \times \dot{M})_{RQ}$.

The case of the PRS is more subtle. Indeed, since the BHs in the PRS quasars are a few times more massive than those in the WRS, for comparable accretion rates or radiative efficiencies we would thus expect $\Gamma$ to be a few times lower in the PRS than in the WRS. Quantitatively, this is very close to what we observe. Therefore, our analysis suggests that $(\eta \times \dot{M})_{PRS} \sim (\eta \times \dot{M})_{WRS}$.

In general, therefore we can conclude that the BHs in the RL quasars have higher accretion rates and/or radiative efficiencies than in the RQ quasars, that is, $(\eta \times \dot{M})_{RL} > (\eta \times \dot{M})_{RQ}$.

What does this mean in terms of the difference of masses between the PRS and WRS? According to equation (12), unless the radiative efficiency parameter is unusually high, the dominant term is the mass accretion rate, and thus we find that $\dot{M}_{BH} \propto \dot{M}$. On the other hand, although the WRS emit more energy than the RQ quasars, they have comparable BH masses, which suggests that the radiative efficiency must also be higher in them. Thus, both $\eta$ and $\dot{M}$ must be higher in RL quasars to achieve $(\eta \times \dot{M})_{PRS} \sim (\eta \times \dot{M})_{WRS}$. Also, because the mass does not increase in the WRS compared to the RQ quasars, while it increases significantly in the PRS quasars, this indicates that the difference of BH masses depends on the difference of radio efficiency, which implies that the two phenomena must be physically coupled, as we observed in Section 3.2.

However, there is one dilemma: assuming $\eta$ is coupled to $\dot{M}$, the fact that $(\eta \times \dot{M})_{PRS} \sim (\eta \times \dot{M})_{WRS}$ does not allow to explain the differences of BH masses and radio luminosities between the PRS and WRS quasars. One way to resolve this problem is to assume in the PRS quasars higher accretion rates and radiative efficiencies, that is, $(\eta \times \dot{M})_{PRS} > (\eta \times \dot{M})_{WRS}$, for a short period of time, sometime in their past. But, it seems that we would also need to assume that such increment in the product $(\eta \times \dot{M})$ produces a SED that is amplified in the radio bands. Interestingly, the model proposed by Sikora & Begelman (2013) already introduced such a distinction in the SED, in terms of the ratio of radio jet efficiency to radiative efficiency. There authors also state that this ratio is reflected in the parameter $R_i$, explaining the bimodality distribution of the RQ and RL quasars. Although we found no such bimodality, we do agree that this model would fit relatively well the separations between the WRS and PRS in $R_i$.

But what about the increase of BH masses? As we mentioned before, the fact that the BHs are more massive only in the PRS suggests that the two phenomena, namely the increase in mass, $\dot{M}_{BH} \propto \dot{M}$, and the increase in radio emission, are coupled. Then the only possibility seems to be that the increase in mass of the BHs in the PRS happens at the same time as their extended radio structures form. An increment by a few tens, or even a hundred, of the product $\eta \times \dot{M}$, during the short timescale necessary to develop the extended radio structures, typically a few $10^8$ yrs, would be sufficient, in good agreement with what is observed in some quasars at high redshifts (e.g., Shemmer et al. 2004; Neri-Larios et al. 2011).

Based on King & Pringle's analysis (King & Pringle 2006), we can test the viability of the hypothesis of a rapid grow in mass of the BHs during the formation of their extended radio structures. Assuming $L_{bol}/L_{Edd} = 1$, we can combine equation (11) and equation (12) and thus integrate to get the solution (Equation (6) in King & Pringle 2006):

$$M/M_0 = \exp[(1/\eta - 1) \, t/t_{Edd}] \quad (14)$$

As we mentioned in Section 2.4, $t_{Edd}$ is comparable to the formation timescale of the extended radio structures. This means that we can put in equation (14) $t/t_{Edd} \sim 1$. During this time the mass of the BH in the PRS would need to grow by a factor three, which means that we can also put $M/M_0 \sim 3$. This yields $\eta \sim 0.47$, which corresponds to an increase by a factor $\sim 5$ of the radiative efficiency (assuming 0.1 as the standard). Obviously, lower radiative accretion efficiencies would produce even higher BH masses, in possibly smaller amounts of time. This shows that a short-lasting high activity phase in the past of the PRS during which $(\eta \times \dot{M})_{PRS} > (\eta \times \dot{M})_{WRS}$, and during which the BH rapidly increases its mass as the extended radio structures form is a highly viable possibility.

Alternatively, assuming $(\eta \times \dot{M})_{PRS} \sim (\eta \times \dot{M})_{WRS}$, could the BH formation timescale be longer in the PRS than in the WRS? As we already indicated, assuming longer timescales for the formation of BHs in the PRS would not explain why we see higher BH masses only in the PRS. This is because in low-density environments, extended radio structures grow over short timescales. So, in the case of substantial timescale differences, there would be no reason to expect the most massive BHs to be connected with such extended radio structures. It seems, therefore, that one would then need to assume higher accretion rates and radiative efficiencies in the PRS than in the WRS, which is in contradiction with the starting hypothesis that they are equal.

Maybe the difference in star formation between the PRS and WRS( cf, Fig. 16c and d), could be interpreted as evidence of an age difference between their galaxy hosts? This would imply that the galaxies hosting the WRS are "younger" than those hosting the PRS. However, our analysis of the galaxy morphologies and their environments show no evidence supporting such assumption. Moreover, we would have the same problem as before with the evolutionary sequence, implying that we observe preferentially quasars in young galaxies. Such conclusion is not supported by our data, because the RQ quasars in Fig. 16b show objects in both SF zone.

Moreover, there are other ways to explain differences of star formation activity in AGNs, like, for example, star bursts triggered by jets in the WRS, or quenching of star formation due to the formation of extended structures in the PRS. However, neither of these alternatives seem fully satisfactory. For example, why only the WRS, which show no evidence of radio jets, would be in a high phase of star formation, while those with apparently strong jets would show quenching? In the absence of generic processes, what physical reasons would explain the different effects the AGNs have on star formation in the PRS and WRS? And, what would be the connection between these different effects and the different BH masses? Therefore, whatever the reasons for the difference of star formation activity between the WRS and PRS, this phenomenon probably cannot explain their other differences. The reverse line of thought would seem the correct way to proceed, first explaining the differences of radio emission and BH masses, then the difference of star formation activity.

In the absence of differences in galaxy host morphologies, in interaction and environments between the PRS and WRS, the hypothesis of transient higher accretion rates and/or radiative efficiencies in the past of the PRS quasars looks, consequently, more probable. We therefore suggest the following scenario. The phenomenon that explain the differences between the RQ and RL, and between the WRS and PRS, must be related to the dynamical behavior of the material falling on the accretion disc (e.g., Tchekhovskoy et al. 2010; Sikora & Begelman 2013; Kellermann et al. 2016). However, as for the specific event that sparks the radio phase, we propose it is stochastic by nature, consistent with a chaotic critical phenomenon.

To become a RL quasar the accretion process must reach a critical state (e.g., Coleman & Dopita 1992; Sikora & Begelman 2013), which happens only rarely, explaining why most quasars are RQ. Even once this critical state is reached, the mass accretion rate does not change much, while the radiative efficiency increases. However, this does not imply a big increment in radio emission efficiency (or radio jet efficiency), and the only structure developing is a radio core. This characterizes the WRS. On the other hand, because this is a chaotic process, the radio emission efficiency can suddenly become extreme, and large-scale radio structures would then develop over very short periods of time (a few $10^8$ yrs). By coupling this increase in radio emission efficiency to a significant increase in mass accretion rate, consistent with the coupling between radio morphologies and optical spectral characteristics we encountered (also, as previously observed by Zamfir, Sulentic & Marziani 2008), the BHs during the radio phase in the PRS would thus grow rapidly, explaining their masses higher by a factor three than in the WRS.

Apparently, there are no contradictions in this scenario. For instance, the differences in BH masses, as we noted be-fore, are relatively small, and since PRS are rare, this would have no impact on the empirical relation between the bulge mass and BH mass (a Magorrian-like diagram), considering typical ranges in mass of the bulges of late and early type galaxies. Therefore, one would not expect any observable difference in galaxy host morphologies between the WRS and PRS. As for the difference of star formation, interactions have nothing to do, since evidence of interactions is as common in the RQ than in the RL quasars.

On the other hand, the chaotic scenario might offer a simple explanation for the lower star formation in the PRS. In general, quasars have a high quantity of gas in their centres, which characterizes their state as quasars, but which may also favour high levels of star formation. Therefore, in principle, we would not need to assume any special mechanism for the high star formation in the RL quasars. This is supported by the fact that many RQ quasars in Fig. 16b are found in the same high SF zone as the WRS quasars. However, since we have a higher number of ionizing photons in the RL quasars than in the RQ quasars, and $(\eta \times \dot{M})_{\mathrm{RL}} > (\eta \times \dot{M})_{\mathrm{RQ}}$, we cannot reject completely the possibility of star formation enhancement in the RL quasars. But in the case of the PRS we also have $(\eta \times \dot{M})_{\mathrm{PRS}} > (\eta \times \dot{M})_{\mathrm{WRS}}$, during a short period of time, which is accompanied by a rapid increase of the masses of the BHs. This implies that more gas in the PRS than in the WRS has fallen onto the BHs, making it unavailable for star formation. A difference of gas mass by a few $10^8 M_\odot$, as we observed, could make a significant difference in terms of star formation, and yet have no impact on the morphology of the hosts if it happens over a short period of time.

## 5 SUMMARY AND CONCLUSION

Comparing the characteristics of a sample of 1958 nearby quasars in the nearby Universe, we found the following:

- RL quasars, with $\log(L_{1.4\mathrm{GHz}}) \geq 22.5$ W Hz$^{-1}$, form $\sim 22\%$ of the quasar population at low redshift, and the majority are WRS, that is, they show only a radio core, with a median LLS of 8 kpc (a mean of 11 kpc), and a radio luminosity $\log(L_{1.4\mathrm{GHz}}) < 24.5$ W Hz$^{-1}$.

- Forming only $\sim 3\%$ of the quasar population, PRS, with $\log(L_{1.4\mathrm{GHz}}) \geq 24.5$ W Hz$^{-1}$, are rare, extended radio sources, with median LLS of 348 kpc, which is about twice the typical value for radio galaxies (e.g., Best et al. 2005a; Lin et al. 2010).

- RQ with $\log(L_{1.4\mathrm{GHz}}) < 22.5$ W Hz$^{-1}$ are not radio dead, but seem to form a continuous distribution in radio luminosity with the RL quasars.

- RQ and WRS quasars have comparable BH masses, while in the PRS BHs are a few times more massive than in the WRS.

- RL (both WRS and PRS) produce more energy in radio, optical and produce more ionizing photons than RQ quasars. This is explained by higher accretion rates and/or radiative efficiencies in RL than in RQ quasars.

- WRS and PRS have comparable accretion rates and/or radiative efficiencies, while differing by two orders of magnitude in radio luminosity.

- There seems to be a coupling between the optical spec-

tral characteristics of the quasars and their radio morphologies: WRS show higher FWHM$_{[OIII]}$ than PRS.

• There is no evidence of a difference in galaxy host morphology, level of interaction and environment between the RQ and RL, or between the WRS and PRS quasars.

• We find no evidence of a difference in star formation in the galaxy hosts of the RQ and RL quasars. However, the PRS seem to have lower levels of star formation than the WRS.

Based on these results, we conclude that the difference between the RQ and RL quasars is intrinsic to the AGN (Kellermann et al. 2016), depending on the dynamical behavior of the falling material or the structure of the accretion disc forming around a BH (e.g., Tchekhovskoy et al. 2010; Sikora & Begelman 2013). However, in the absence of evidence of generic processes, like galaxy formation or interaction, we propose that the event that triggers the radio-loud phase is stochastic by nature, consistent with a chaotic critical phenomenon.

Most quasars in the Universe are RQ or WRS, because the normal accretion rate and radiative efficiency of BHs is low (see also Sikora & Begelman 2013). Only a few RL quasars exceed a critical level necessary to develop an extended radio structure typical of PRS. Once a quasar reaches the critical limit, the radio structure expands through space in a burst-like manner, and a temporary increase in the mass accretion rate during this radio-loud phase explains why BHs are more massive in the PRS than in the RQ and WRS quasars.

Note that according to this scheme, recurrent radio bursts in PRS are possible, assuming the first burst does not empty the gas reservoir completely near the BH (Heckman & Best 2014). Additionally, we might expect the timescale for the radio activity to significantly decrease as the radio emission efficiency increases. Consequently, different evolution patterns for the co-moving number densities of PRS and WRS quasars in the Universe could be expected, possibly similar to what was observed for narrow-line radio galaxies (e.g., Donoso, Best & Kauffmann 2009).

## ACKNOWLEDGEMENTS

The authors acknowledge the comments and suggestions made by Sam Lindsay, assistant editor of MNRAS, and the comments of an anonymous referee, which help us improving the quality of our analysis and presentation. This study was possible thanks to DAIP-Ugto (1006/2016). RAOM acknowledges support from CONACYT (Mexico) through PhD fellowship 322095, and CAPES (Brazil) through a PDJ fellowship from project 88881.030413/2013-01, program CSF-PVE. The Funding for SDSS has been provided by the Alfred P. Sloan Foundation, the Participating Institutions, the National Science Foundation, and the U.S. Department of Energy Office of Science. The full acknowledgement can be found here: www.sdss3.org. This research has made use of the R language and environment for statistical computing, available at www.r-project.org. This paper also made use of the VizieR catalog access tool, CDS, Strasbourg (Ochsenbein et al. 2002) and of the Wide-field Infrared Survey Explorer, which is a joint project of the University of California, Los Angeles, and the Jet Propulsion Laboratory/California Institute of Technology, funded by the National Aeronautics and Space Administration.

This paper has been typeset from a TEX/LATEX file prepared by the author.